\documentclass[12pt]{article}
\usepackage{epsfig,amsmath,amssymb,latexsym,graphicx,setspace,fullpage}

\def\bib{\vskip12pt\par\noindent\hangindent=0.5 true cm\hangafter=1}

\newcommand{\by}{\mbox{\boldmath$y$}}
\newcommand{\ba}{\mbox{\boldmath$a$}}

\newcommand{\bx}{\mbox{\boldmath$x$}}
\newcommand{\bX}{\mbox{\boldmath$X$}}
\newcommand{\bY}{\mbox{\boldmath$Y$}}
\newcommand{\bI}{\mbox{\boldmath$I$}}

\newcommand{\bu}{\mbox{\boldmath$u$}}
\newcommand{\bv}{\mbox{\boldmath$v$}}
\newcommand{\balpha}{\mbox{\boldmath$\alpha$}}
\newcommand{\bbeta}{\mbox{\boldmath$\beta$}}
\newcommand{\bfeta}{\mbox{\boldmath$\eta$}}
\newcommand{\bepsilon}{\mbox{\boldmath$\epsilon$}}
\newcommand{\bgamma}{\mbox{\boldmath$\gamma$}}
\newcommand{\bmu}{\mbox{\boldmath$\mu$}}
\newcommand{\bSigma}{\mbox{\boldmath$\Sigma$}}

\newcommand{\btheta}{\mbox{\boldmath$\theta$}}
\def\sn{\sqrt{n}}

\def\mA{\mathcal{A}}
\def\mN{\mathcal{N}}
\def\mO{\mathcal{O}}

\begin{document}

{\baselineskip=18pt
{
\vspace*{10mm}
{\renewcommand{\thefootnote}{\fnsymbol{footnote}}
\begin{center}
{\LARGE
Bayesian projection approaches to variable selection } \\

\vspace{7mm}
{\LARGE and exploring model uncertainty} \\
 
 \vspace*{15mm}
 
{\Large David J. Nott and Chenlei Leng\footnote{David
J. Nott is Associate Professor, Department of Statistics and Applied Probability,
National University of Singapore, Singapore 117546. (email
standj@nus.edu.sg).  Chenlei Leng is Assistant Professor, Department of Statistics
and Applied Probability, National University of Singapore, 
Singapore 117546 (email stalc@nus.edu.sg).  This work was partially supported
by an Australian Research Council grant.}}

\end{center}

\doublespacing

\vspace*{4mm}
\begin{abstract}
A Bayesian approach to variable selection which is based on the expected Kullback-Leibler divergence 
between the full model and its projection onto a submodel has recently been suggested in the literature.  
Here we extend this idea by considering projections onto subspaces defined via some form of $L_1$ constraint 
on the parameter in the full model.  This leads to Bayesian model selection approaches related to 
the lasso.  In the posterior distribution of the projection there is positive probability that 
some components are exactly zero and the posterior distribution on the model space induced by 
the projection allows exploration of model uncertainty.  We also consider use of the approach 
in structured variable selection problems such as ANOVA models where it is desired to incorporate 
main effects in the presence of interactions.  Here we make use of projections related to the non-negative garotte
which are able to respect the hierarchical constraints.  We also prove a consistency result 
concerning the posterior distribution on the model induced by the projection, and show that 
for some projections related to the adaptive lasso and non-negative garotte the posterior 
distribution concentrates on the true model asymptotically.  
\end{abstract}

\vspace*{5mm}\noindent
{\it Keywords}:  
Bayesian variable selection, Kullback-Leibler projection, lasso, non-negative garotte, preconditioning.   

\section{Introduction}

Bayesian approaches to model selection and describing model uncertainty have become 
increasingly popular in recent years.  In this paper we extend a method of variable selection
considered by Dupuis and Robert (2003) and related to earlier suggestions by Goutis
and Robert (1998) and Mengersen and Robert (1996).  Dupuis and Robert (2003) consider
an approach to variable selection where models are selected according to a relative
explanatory power, where relative explanatory power is defined using the expected
Kullback-Leibler divergence between the full model and its projection onto 
a submodel.  The goal is to find the most parsimonious model which achieves an acceptable
loss of explanatory power compared to the full model.  

In this paper we consider an extension of the method of Dupuis and Robert (2003) where
instead of considering a projection onto a subspace defined by a set of active covariates
we consider projection onto a subspace defined by some other form of constraint on the parameter in
the full model.  Certain choices of the constraint (an $L_1$ constraint as used in the
lasso of Tibshirani (1996), for example) lead to exact zeros for some of the coefficients.  
The kinds of projections we consider also have computational advantages, in that parsimony is
controlled by a single continuous parameter and we avoid the search
over a large and complex model space.  Searching the model space in traditional Bayesian 
model selection approaches with large numbers of covariates is a computationally daunting
task.  In contrast, with our method we handle model uncertainty in a continuous way through an encompassing
model, and can exploit existing fast lasso type algorithms to calculate
projections for samples from the posterior distribution in the encompassing model.  This
allows sparsity and exploration of model uncertainty while preserving approximately 
posterior predictive behaviour based on the full model.  Furthermore, the method is easy to implement
given a combination of existing Bayesian software and software implementing lasso type fitting methods.  
While the method is more computationally intensive than calculating a solution path for
a classical shrinkage approach like the lasso, this is the price to be paid for exploring
model uncertainty and the computational demands of the method are certainly less than Bayesian
approaches which search the model space directly.  Our idea can
also be applied to structured variable selection problems such as those arising in ANOVA models
where we might wish to include interaction terms only in the presence of the corresponding
main effects.  A plug-in
version of our approach is also related to the preconditioning method of Paul {\it et al.} (2007)
for feature selection in ``large $p$, small $n$" regression problems.  
A key advantage of our approach is simplicity in prior specification.  
One is only required to specify a prior on the parameter in the full model, and not a prior
on the model space or a prior on parameters for every submodel.  Nevertheless, the posterior
distribution on the model space induced by the projection can be used in a similar
way to the posterior distribution on the model space in a traditional Bayesian analysis 
for exploring model uncertainty and different interpretations of the data.  

There are many alternative Bayesian strategies for model selection and exploring
model uncertainty to the one considered here.  Bayes factors and Bayesian model
averaging (Kass and Raftery, 1995, Hoeting {\it et al.}, 1999, Fern\'{a}ndez {\it et al.}, 2001) are the traditional
approaches to addressing issues of model uncertainty in a Bayesian framework.  
As already mentioned, prior specification can be very demanding for these approaches, although
general default prior specifications have been suggested (Berger and Pericchi, 1996;
O'Hagan, 1995).  In our later examples we focus on generalized linear models, and 
Raftery (1996) suggests some reference priors for Bayesian model comparison in this
context.  Structuring priors hierarchically and estimating hyperparameters
in a data driven way is another way to reduce the complexity of prior specification, 
and this can work well (George and Foster, 2000).  Various Bayesian predictive criteria for selection
have also been suggested (Laud and Ibrahim, 1995, Gelfand and Ghosh, 1998, 
Spiegelhalter {\it et al.}, 2002).  These approaches generally do not require specification
of a prior on the model -- however, prior specification for parameters in all models is
still required and this can be quite demanding if there are a large number of models to be compared.  
Decision theoretic strategies which attempt to take account of the costs of data collection
for covariates have also been considered (Lindley, 1968, Brown {\it et al.}, 1999, Draper and Fouskakis, 2000).  
Bayesian model averaging can also be combined with model selection as in Brown
{\it et al.} (2002).  The projection method of Dupuis and Robert (2003) that we extend here is related to
the Bayesian reference testing approach of Bernardo and Rueda (2002) and the predictive method of
Vehtari and Lampinen (2004). There are also less formal approaches to Bayesian model
comparison including posterior predictive
checks (Gelman, Meng and Stern, 1996) which are targeted according to the uses that will be
made of a model.  We see one application of the methods we describe here as being to suggest a small
set of candidate simplifications of the full model which can be examined by such means as to
their adequacy for specific purposes.  

The projection methods we use here for model selection are related to the lasso of
Tibshirani (1996) and its many later extensions.  There has been some recent work on 
incorporating the lasso into Bayesian approaches
to model selection.  Tibshirani (1996) pointed out the Bayesian interpretation of the lasso
as a posterior mode estimate in a model with independent double exponential priors
on regression coefficients.  Park and Casella (2008) consider the Bayesian lasso, where 
estimators other than the posterior mode are considered -- their estimators do not provide
automatic variable selection via the posterior mode but convenient computation and inference are possible within
their framework.  Yuan and Lin (2005) consider a hierarchical prior formulation in Bayesian
model comparison and a certain analytical approximation to posterior probabilities connecting
the lasso with the Bayes estimate.  Recently Griffin and Brown (2007) have also
considered alternatives to double exponential prior distributions on the coefficients to provide
selection approaches related to the adaptive lasso of Zou (2006).   

The structure of the paper is as follows.  In the next section we briefly review the method of
Dupuis and Robert (2003) and consider our extension of their approach.  Computational
issues and predictive inference are considered in Section 3, and then 
a consistency result relating to the posterior distribution on model space
induced by the projection is proved in Section 4.   We describe applications to structured variable
selection in Section 5 and Section 6 considers connections between our method and the
preconditioning approach to selection of Paul {\it et al.} (2007) in ``large $p$, small $n$" 
regression problems.  Section 7 considers some examples and simulation studies and
Section 8 concludes.  

\section{Projection approaches to model selection}

\subsection{Method of Dupuis and Robert}

Dupuis and Robert (2003) consider a method of model selection based on the Kullback-Leibler 
divergence between the true model and its projection onto a submodel.  Suppose we are considering
a problem of variable selection in regression, where $M_F$ denotes the full model including
all covariates and $M_S$ is a submodel with a reduced set of covariates.  
Write $f(\by|\btheta_F,M_F)$ and $f(\by|\btheta_S,M_S)$
for the corresponding likelihoods of the models with parameters $\btheta_F$ and $\btheta_S$.  Let $\btheta_S'=\btheta_S'(\btheta_F)$ be the projection of $\btheta_F$
onto the submodel $M_S$.  That is, $\btheta_S'$ is the value for $\btheta_S$ for which
$f(\by|\btheta_S,M_S)$ is closest in Kullback-Leibler divergence to $f(\by|\btheta_F,M_F)$, so that
$$\btheta_S'=\mbox{arg}\min_{\btheta_S}\int \log\frac{f(\bx|\btheta_F,M_F)}{f(\bx|\btheta_S,M_S)}f(\bx|\btheta_F,M_F)d\bx.$$
Let
$$\delta(M_S,M_F)=\int \int \log\frac{f(\bx|\btheta_F,M_F)}{f(\bx|\btheta_S',M_S)}f(\bx|\btheta_F,M_F)d\bx\,\, p(\btheta_F|y)d\btheta_F$$
be the posterior expected Kullback-Leibler divergence between the full model and its Kullback-Leibler
projection onto the submodel $M_S$.  The relative loss of explanatory power for $M_S$ is
$$d(M_S,M_F)=\frac{\delta(M_S,M_F)}{\delta(M_0,M_F)}$$
where $M_0$ denotes the model with no covariates and Dupuis and Robert (2003) suggest model selection by choosing the subset model most
parsimonious for which $d(M_S,M_F)<c$ where $c$ is an appropriately small constant.  
If there is more than one model of the minimal size satisfying the bound, then the one with
the smallest value of $\delta(M_S,M_F)$ is chosen.  Dupuis and Robert (2003) show that
$\delta(M_0,M_F)$ can be interpreted as
measuring the explanatory power of the full model, and using an additivity property
of projections they show that $d(M_S,M_F)<c$ guarantees that
our chosen submodel $S$ has explanatory power at least $100(1-c)\%$ of the explanatory
power of the full model.  This interpretation is helpful in choosing $c$.  
For a predictive quantity $\Delta$ we further suggest approximating the predictive density
$p(\Delta|\by)$ for a chosen subset model $S$ by 
\begin{eqnarray}
 p_S(\Delta|\by)  & = & \int p(\Delta|\theta_S',\by)p(\btheta_S'|\by)d\btheta_S'  \label{predictive}
\end{eqnarray}
where $p(\btheta_S'|\by)$ is the posterior distribution of $\btheta_S'$ under
the posterior distribution of $\btheta_F$ for the full model. 

\subsection{Extension of the method}

To be concrete suppose we are considering variable selection for generalized linear
models.  Write $y_1,...,y_n$ for the responses with $E(y_i)=\mu_i$ and suppose
that each $y_i$ has a distribution from the exponential family
$$f\left(y_i;\theta_i,\frac{\phi}{A_i}\right)=\exp\left(\frac{y_i\theta_i-b(\theta_i)}{\phi/A_i}+c\left(y_i;\frac{\phi}{A_i}\right)\right)$$
where $\theta_i=\theta_i(\mu_i)$ is the natural parameter, $\phi$ is a scale parameter, the $A_i$
are known weights and $b(\cdot)$ and $c(\cdot)$ are known functions.  For a smooth
invertible link function $g(\cdot)$ we have $\eta_i=g(\mu_i)=\bx_i^T\bbeta$ where $\bx_i$ is a 
$p$-vector of covariates and $\bbeta$ is a $p$-dimensional parameter vector.  Writing $\bX$ for
the design matrix with $i$th row $\bx_i^T$ and $\bfeta=(\eta_1,...,\eta_n)^T$ (the values of
$\bfeta$ are called the linear predictor values) we have 
$\bfeta=\bX\bbeta$.  We write $f(\by;\bbeta)$ for the likelihood.  Now let $\bbeta$ be fixed and suppose
we wish to find for some subspace $S$ of the parameter space the Kullback-Leibler projection
onto $S$.  The subspace $S$ might be defined by a subset of ``active" covariates as in
Dupuis and Robert (2003) but here we consider subspaces such as
\begin{eqnarray}
 S & = & S(\lambda)=\left\{\bbeta:  \sum_{j=1}^p |\beta_j|\leq \lambda\right\} \label{subspace1}
\end{eqnarray}
\begin{eqnarray}
 S & = & S(\bbeta^*,\lambda)=\left\{\bbeta: \sum_{j=1}^p |\beta_j|/|\beta_j^*|\leq \lambda\right\} \label{subspace1.5}
\end{eqnarray}
where $\bbeta^*$ is a parameter value that supplies weighting factors in the constraint or
\begin{eqnarray}
 S & = & S(\lambda,\eta)=\left\{\bbeta: \sum_{j=1}^p |\beta_j|+\eta \sum_{j=1}^p {\beta_j}^2\leq \lambda\right\}.
 \label{subspace2}
\end{eqnarray}
The choice (\ref{subspace1}) leads to procedures related to the lasso of Tibshirani (1996), 
(\ref{subspace1.5}) relates to the adaptive lasso of Zou (2006) and 
(\ref{subspace2}) is related to the elastic net of Zou and Hastie (2005).  
In (\ref{subspace1.5}) we have allowed the space that we are projecting onto to depend
on some parameter $\bbeta^*$, and later we will allow $\bbeta^*$ to be the parameter
in the full model that we are projecting, so that the subspace that we are projecting onto
is adapting with the parameter.  There is no reason to forbid this in what follows.  
In a later section we also consider projections which are related to Breiman's (1995) non-negative garotte
and which allow for structured variable selection in the presence of hierarchical relationships among
predictors.  The close connection between the adaptive lasso and the non-negative garotte is
discussed in Zou (2006).  

Now let $\bbeta_S$ be a parameter in the subspace $S$.  In the development below
we consider the scale parameter $\phi$ as known -- we consider an unknown scale
parameter later.  The Kullback-Leibler divergence
between $f(\by;\bbeta)$ and $f(\by;\bbeta_S)$ is 
\begin{eqnarray}
& & E_{\bbeta}\left(\log \frac{f(\bY;\bbeta)}{f(\bY;\bbeta_S)}\right) \label{kld}
\end{eqnarray}
where $E_{\bbeta}$ denotes the expectation with respect to $f(\by;\bbeta)$.  Writing $\mu_i(\bbeta)$
and $\theta_i(\bbeta)$ for the mean and natural parameter for $y_i$ when the parameter is $\bbeta$ 
(\ref{kld}) is given by
\begin{eqnarray*}
 & & E_{\bbeta}\left(\sum_{i=1}^n \frac{Y_i\theta_i(\bbeta)-b(\theta_i(\bbeta))}{\phi/A_i} - \sum_{i=1}^n \frac{Y_i\theta_i(\bbeta_S)-b(\theta_i(\bbeta_S))}{\phi/A_i}\right) \\
 & = & \sum_{i=1}^n \frac{\mu_i(\bbeta)\theta_i(\bbeta)-b(\theta_i(\bbeta))}{\phi/A_i}-\sum_{i=1}^n
          \frac{\mu_i(\bbeta)\theta_i(\bbeta_S)-b(\theta_i(\bbeta_S))}{\phi/A_i} \\
  & = & -\log f(\mu(\bbeta);\bbeta_S)+C
\end{eqnarray*}
where $f(\mu(\bbeta);\bbeta_S)$ is the likelihood evaluated at $\bbeta_S$ with data $\by$ replaced
by fitted means $\bmu(\bbeta)$ when the parameter is $\bbeta$ and $C$ represents terms not
depending on $\bbeta_S$ and hence irrelevant when minimizing over $\bbeta_S$.  
So minimization with respect to $\bbeta_S$ subject to a constraint just 
corresponds to minimization of the negative log-likelihood subject to a constraint
but with data $\bmu(\bbeta)$ instead of $\by$.  
Dupuis and Robert (2003) observed that in the case where the subspace $S$ is defined by
a set of active covariates calculation of the Kullback-Leibler projection can be done using standard
software for calculation of the maximum likelihood estimator in generalized linear models:  one
simply ``fits to the fit" using the fitted values for the full model instead of the responses $\by$
in the fitting for a subset model.  Clearly for some choices of the response distribution
the data $\by$ might be integer valued, 
but commonly generalized linear modelling software does
not check this condition so that replacement of the data $\by$ with the fitted means 
for the full model can usually be done.  

In our case, suppose we wish to calculate the projection onto the subspace (\ref{subspace1}).  
We must minimize 
$$-\log f(\bmu(\bbeta);\bbeta_S)\mbox{    subject to    }\sum_{j=1}^p |\bbeta_{S,j}|\leq \lambda$$
where $\beta_{S,j}$ denotes the $j$th element of $\bbeta_S$.  This is equivalent to minimization of
$$-\log f(\bmu(\bbeta);\bbeta_S)+\delta \sum_{j=1}^p |\beta_{S,j}|$$
for some $\delta>0$.  Here the calculation just involves the use of the lasso 
of Tibshirani (1996) where in the calculation the responses are replaced by the
fitted values $\bmu(\bbeta)$.  In the case of a Gaussian linear model, 
the whole solution path over values of $\delta$ can be
calculated with computational effort equivalent to a single least squares fit 
(Osborne {\it et al.}, 2000, Efron {\it et al.}, 2004).  Efficient algorithms
are also available for generalized linear models (Park and Hastie, 2007).  Note that
because the relationship between $\lambda$ and $\delta$ depends on $\bbeta$, generally we
calculate the whole solution path over $\delta$ in order to calculate the projection
onto the subscpace defined by the constraint.  
One can consider other constraints apart from an $L_1$ constraint.    
For instance, (\ref{subspace1.5}) leads to minimization of
$$-\log f(\bmu(\bbeta);\bbeta_S)+\delta\sum_{j=1}^p |\beta_{S,j}|/|\beta_j|$$
for $\delta>0$ if we choose $\bbeta^*=\bbeta$ which gives a certain adaptive lasso estimator 
(Zou, 2006) obtained by fitting with data $\by$ replaced by $\bmu(\bbeta)$.  
Also, (\ref{subspace2}) leads to minimization of 
$$-\log f(\bmu(\bbeta);\bbeta_S)+\delta \sum_{j=1}^p |\beta_{S,,j}|+\gamma \sum_{j=1}^p \beta_{S,j}^2$$
for positive constants $\delta$ and $\gamma$ which is related to the elastic
net of Zou and Hastie (2005).  

Later we focus on the lasso and adaptive lasso type projections for which the parameter $\lambda$ needs to
be chosen.  One way to do this is to follow a similar strategy to the one employed in Dupuis and
Robert (2003).  Writing $M_S=M_S(\lambda)$ for the model subject
to the restriction (\ref{subspace1}) or (\ref{subspace1.5}), we choose $\lambda$ as small as possible subject to
$d(M_S,M_F)<c$.  Note that choosing the single parameter $\lambda$ is much easier
than searching over subsets as in Dupuis and Robert (2003), and that $d(M_S,M_F)$ 
increases monotonically as $\lambda$ decreases.  An alternative to choosing $c$ based
on relative explanatory power would be to directly choose the observed sparsity in
the model:  that is, to choose $\lambda$ so that the posterior mean of the number of
active components in the projection is equal to some specified value.  The relative
loss of explanatory power can also be reported for this choice.  Another possibility is to avoid
choosing $\lambda$ at all, but instead to simply report the characteristics of the models appearing
on the solution path over different samples from the posterior distribution in the full model.  

\subsection{Gaussian response with unknown variance}

So far we have considered the scale parameter $\phi$ to be known.  For binomial and
Poisson responses $\phi=1$, but we also wish to consider Gaussian linear models with
unknown variance $\phi=\sigma^2$.  Calculation of projections is still straightforward in
the Gaussian linear model with unknown variance parameter. Now suppose we have
mean and variance parameter $\bbeta$ and $\sigma^2$, and write $\bbeta_S$ and 
$\sigma_S^2$ for corresponding parameter values in some subspace $S$.  In this case (\ref{kld}) is
\begin{eqnarray*}
& &  E_{\bbeta,\sigma^2}\left(-\frac{n}{2}\log 2\pi\sigma^2-\sum_{i=1}^n \frac{(Y_i-\mu_i(\bbeta))^2}{2\sigma^2}
      +\frac{n}{2}\log(2\pi\sigma_S^2)+\sum_{i=1}^n \frac{(Y_i-\mu_i(\bbeta_S))^2}{2\sigma_S^2}\right) \\
  & = & -\frac{n}{2}\log2\pi\sigma^2-\frac{n}{2}+\frac{n}{2}\log2\pi\sigma_S^2 
    +\frac{1}{2\sigma_S^2}\sum_{i=1}^n (\sigma^2+(\mu_i(\bbeta)-\mu_i(\bbeta_S))^2) \\
  & = & \frac{n}{2}\left(\log\frac{\sigma_S^2}{\sigma^2}-1\right)+\frac{n\sigma^2}{2\sigma_S^2}+
    \frac{1}{2\sigma_S^2}\sum_{i=1}^n (\mu_i(\bbeta)-\mu_i(\bbeta_S))^2.
\end{eqnarray*}
Minimization with respect to $\bbeta_S$ subject to an $L_1$ constraint involves minimization
of 
$$\sum_{i=1}^n (\mu_i(\bbeta)-\mu_i(\bbeta_S))^2+\delta\sum_{j=1}^p |\beta_{S,j}|$$
and the minimizer $\bbeta_S'$ over 
$\bbeta_S$ is independent of the value of
$\sigma_S^2$.  Once the projection $\bbeta_S'$ is calculated, the projection 
${\sigma_S^2}'$ is easily shown from the expression above to be
$${\sigma_S^2}'=\sigma^2+\frac{(\bmu(\bbeta)-\bmu(\bbeta_S'))^T(\bmu(\bbeta)-\bmu(\bbeta_S')}{n}.$$

\section{Computation and predictive inference}

We have already discussed computation of the Kullback-Leibler projection onto subspaces 
of certain forms in generalized linear models.  Hence generating from the posterior distribution 
of the projection is easily done -- we simply generate a sample from the posterior distribution
$p(\bbeta|\by)$ of the parameter, $\bbeta^{(1)},...,\bbeta^{(s)}$ say, and then for each of these parameter values
we calculate the corresponding projections ${\bbeta^{(1)}}',...,{\bbeta^{(s)}}'$.  
Note that the pattern of sparsity of the projection is different for different samples
from the posterior distribution, so that the posterior distribution of the projection
provides one way of exploring model uncertainty.
We approximate the predictive density (\ref{predictive}) by
$$\frac{1}{s}\sum_{i=1}^s p(\Delta|{\bbeta^{(i)}}',\by)$$
where $p(\Delta|{\bbeta^{(i)}}',y)$ denotes the predictive distribution for $\Delta$ given the
parameter value ${\bbeta^{(i)}}'$ and data $\by$.  We can write 
$\gamma_j'=I(\beta_j'\neq 0)$ for the indicator of whether or not the $j$th component
of the projection of $\bbeta$ is nonzero, and $\bgamma'=(\gamma_1',...,\gamma_p')^T$.  
Then we can write $p_S(\Delta|\by)$ in a different form to (\ref{predictive}), namely
$$p_S(\Delta|\by)=\sum_{\bgamma'} p(\bgamma'|\by)p(\Delta|\bgamma',\by)$$
where
$$p(\Delta|\bgamma',\by)=\int p(\Delta|\bgamma',\bbeta',\by)p(\bbeta'|\bgamma',\by)d\bbeta'.$$
These expressions for predictive densities are formally similar to those arising in Bayesian 
model averaging, where
different values for the indicators $\bgamma'$ define different models.  Of course, the posterior
distribution on $\bgamma'$ cannot be interpreted in quite the same way as the posterior
distribution on the model space in a formal Bayesian approach to model comparison, but we still
believe that examining $p(\bgamma'|\by)$ can be helpful for exploring different interpretations
of the data in our approach.  We illustrate this in the examples below.  

\section{Consistent model selection}

Let $\bbeta^0$ denote the true parameter, and for any $\bbeta$ write 
$\mA(\bbeta)=\{k: \beta_k\neq 0\}$ so that for instance $\mA(\bbeta^0)$ is
the set of nonzero coefficients for the true parameter.  Suppose that $\bbeta$ is some
fixed parameter value and consider $\bbeta_S'$ which minimizes
\begin{equation}
-\log p(\bmu(\bbeta); \bbeta_S)~\text{subject to}~
  \sum_{j=1}^p |\beta_{S,j}|/|\beta_j| \le \lambda. \label{obj}
\end{equation}
with respect to $\bbeta_S$.  That is, we consider in this section Kullback-Leibler projections
for subspaces of the form (\ref{subspace1.5}) related to the adaptive lasso of Zou (2006).  
A similar result to the one below can be proved for some projections related to the non-negative
garotte in view of the close connection between the adaptive lasso and the non-negative
garotte (Zou, 2006).  We will examine projections related to the non-negative garotte when
we look at structured variable selection problems later.  

Considering only the case of a generalized linear model, the minimization problem above
is equivalent to minimization of
\begin{eqnarray}
 & & \sum_{i=1}^n \left\{ \mu_i(\bbeta)\theta_i(\bbeta_S)+b(\theta_i(\bbeta_S))\right\} +
\gamma\sum_{j=1}^p |\beta_{S,j}|/|\beta_j| \label{obj2}
\end{eqnarray}
where there is a natural one to one correspondence between $\gamma$ and $\lambda$
in (\ref{obj}) and for simplicity we are considering the case where the observation specific
weights $\phi/A_i$ are all equal.  Actually, as mentioned earlier, the relationship between
$\gamma$ and $\delta$ depends on $\bbeta$, but this can be ignored in what follows:    
if we take $\lambda= \# \mA(\bbeta^0)+O_p(1/\sn)$, where $\#\mA(\bbeta^0)$ is
the number of the entries in $\mA(\bbeta^0)$, the consistency result
for (\ref{obj}) can be similarly established.
We also consider henceforth the natural link function 
$\theta_i(\bbeta)=\bx_i^T\bbeta$, although the arguments below extend easily to other
link functions and the case of unequal weights for the observations.  

If $\bbeta$ is a sample from the posterior distribution then under general conditions
it is a root-$n$ consistent estimator, and
we know that for any $\epsilon\in [0,1]$ there exists $C$ not depending on $n$
such that with $\mN_n=\{\balpha: \|\balpha-\bbeta^0\|\leq C/\sqrt{n}\}$, 
$Pr(\bbeta\in \mN_n)\geq 1-\epsilon$  where the probability is calculated with
respect to the distribution
$$q(\bbeta)=\int p(\bbeta|\by)p(\by|\bbeta^0)dy.$$
We will show that 
$$\lim_{n\rightarrow\infty} Pr(\mA(\bbeta_S')=\mA(\bbeta^0)\mbox{ for every }\bbeta\in\mN_n)=1$$
for a suitable sequence of values $\gamma_n$ for $\gamma$ and hence
$$\lim_{n\rightarrow\infty} Pr(\mA(\bbeta_S')=\mA(\bbeta^0))=1.$$
That is, the posterior distribution of the projection indicates the correct model
with probability one as $n\rightarrow\infty$ for a suitable choice of the sequence
of parameters $\gamma_n$ defining the projection.  

We assume the following regularity conditions, which are the same as in Zou (2006).
\begin{itemize}
\item[1.] The Fisher information $I(\bbeta^0)$ is positive definite;
\item[2.] There is a large enough open set $\mO$ containing the true 
parameter $\bbeta^0$
such that $\forall \bbeta \in \mO$,
\[|b'''(\bx^T \bbeta)| \le M(x) < \infty \] and
\[ E[M(\bx)|x_ix_jx_k|] < \infty \]
for any $i,j,k$.
\end{itemize}

\vspace{0.4cm}
\noindent
{\bf Theorem 1.} For $\bbeta \in \mN_n$, if $\gamma_n/\sn \rightarrow 0$ and
$\gamma_n \rightarrow \infty$ as $n\rightarrow\infty$, $\bbeta_S'$ which minimizes (\ref{obj2})
is consistent in variable selection and is $\sn$-consistent.

\noindent
The proof, which is an easy adaptation of a similar result in Zou (2006), is given in 
the Appendix.

\section{Structured variable selection}

In the last section we considered variable selection in generalized linear models:  as before, 
write $\bfeta=\bX\bbeta$ where $\bfeta$ is the vector of linear predictor values, $\bX$ is a design
matrix and $\bbeta$ is a parameter vector.  In this section we will combine the structured
variable selection approach of Yuan, Joseph and Zou (2007) which uses the non-negative
garotte of Breiman (1995) with our projection approach to variable selection.  

Consider the model in which $\bfeta=X(\bbeta^*\circ \btheta)$ where $\bbeta^*=(\beta^*_1,...,\beta^*_p)^T$
and $\btheta=(\theta_1,...,\theta_p)^T$ are $p$-vectors of parameters with $\theta_j\geq 0$, 
$\sum_{j=1}^p \theta_j\leq p$ and where $\circ$ denotes element by element multiplication
of two vectors.  We write $\bbeta^*$ instead of $\bbeta$ to emphasize that $\bbeta^*$ is a different
parameter in a different model to the original one, although our original model can be recovered
by setting $\bbeta^*=\bbeta$ and $\btheta$ a $p$-vector of ones.  We can consider the projection of this
parameter onto the subspace
$$S=S(\bbeta,\lambda)=\{(\bbeta^*,\theta): \bbeta^*=\bbeta,\sum_{j=1}^p \theta_j\leq \lambda, \theta_j\geq 0, j=1,...,p\}.$$
To calculate the projection we need to minimize 
$-\log p(\bmu(\bbeta); \bbeta\circ \btheta)$ with respect to $\btheta$ subject to $\theta_j\geq 0$ and
$\sum_{j=1}^p \theta_j\leq \lambda$.  For the Gaussian case, this is just Breiman's non-negative
garotte applied to the fitted values $\bmu(\bbeta)$ rather than the data $\by$.  The minimization
problem is easily solved.  See Yuan, Joseph and Zou (2007) for computational details
in the slightly more complicated situation of structured fitting of generalized linear models.  
As pointed out by Zou (2006), the non-negative garotte is very closely related to the
adaptive lasso.  

In solving the minimization problem above we may find that some of the $\theta_j$ are zero.  This
allows variable selection in the original model for which we have replaced the parameter $\bbeta$
with $\bbeta^*\circ \btheta$.  Yuan, Joseph and Zou (2007) also suggested a way in which hierarchical
structure can be incorporated in the non-negative garotte, and we make use of this idea here.  
Following their notation, for the $i$th predictor (corresponding to the $i$th column of $\bX$) we
write ${\cal D}_i$ for the set of predictors which are so-called parents of $i$.  Under the
strong heredity principle (Chipman, 1996) all the predictors in ${\cal D}_i$ must be included in the model before
the $i$th predictor is included.  Under the weak heredity principle at least one of the predictors
in ${\cal D}_i$ must be included before the $i$th predictor is included.  Yuan, Joseph and Zou (2007)
suggest the constraints $\theta_i\leq \theta_j$ for $j\in {\cal D}_i$ to enforce the strong
heredity principle and $\theta_i\leq \sum_{j\in {\cal D}_i}\theta_j$ to enforce the weak heredity
principle.  The linear nature of the constraints ensures that computations are still tractable.   

\section{``Large $p$, small $n$" problems and preconditioning}

Paul {\it et al.} (2007) suggest that in ``large $p$, small $n$" regression problems with more predictors
than observations it is beneficial to separate the problem of obtaining good predictions from that
of variable selection.  With this in mind, they suggest a two step procedure where first a good
predictor $\hat{\by}$ is found for the mean response, and then in a second stage a model
selection and fitting procedure such as the lasso is applied with the responses $\by$ replaced
by $\hat{\by}$.  They show that such a procedure can perform better than application of the fitting and
selection procedure to the raw outcome $\by$.  

We note that their second stage of fitting to a set of fitted values (using the lasso in the case
of a linear model for instance) 
commonly corresponds to calculation of a Kullback-Leibler projection if $\hat{\by}$ is obtained
by plugging in of a point estimate of the model parameters.  Our approach is related though
slightly different -- we generate from the posterior distribution of the parameters, and for each
draw from the posterior we fit to the corresponding set of fitted values.  

\section{Examples and simulations}

\subsection{Low birthweight data}

We consider application of our approach to the low birthweight data of Hosmer and Lemeshow (1989).  
The data are concerned with 189 births at a US hospital.  We consider a logistic
regression model for a response which is a binary indicator
for birthweight being less than 2.5kg.  The predictors in the model are shown in Table 1.  
These predictors had been shown to be associated with low birthweight in past studies and
it was desired to find out which of the predictors were important 
for the medical centre where the data were collected.  In our analysis
we leave the binary predictors unchanged, but centre and scale the other predictors to
have mean zero and variance one.  We fit the full model with a prior on the coefficients that
is normal, with mean vector $0$ and covariance matrix $3\bI$ where $\bI$ denotes the identity
matrix.  This is a fairly noninformative prior on the scale of the probabilities:  note that
making the prior variances of coefficients very large would correspond to a very informative prior
on the probability scale where high prior probability is placed on coefficient values corresponding
to most of the fitted values being close to zero or one.  
Coefficient estimates (posterior means) and posterior standard deviations obtained by 
fitting the full model are shown in Table 2.  These
results were obtained using the MCMCpack package in R (Martin and Quinn, 2007).
To obtain the results reported we ran the MCMC scheme for 1000 ``burn in" and 
10000 sampling iterations.  

We consider our projection approach to selection with the lasso type constraint (\ref{subspace1})
as well as the adaptive lasso type constraint (\ref{subspace1.5}).  
For each sample from the posterior, the whole solution path was calculated for the projection
as the parameter $\lambda$ was varied, and all the distict models on the path were recorded. 
Thus for each sample from the posterior, we should have roughly 11 distinct models 
(including the null and the full models) because there are 10 covariates.  We say ``roughly"
11 distinct models because it is possible for a variable to leave the model as the regularization parameter
is increased in the solution path, but this is not very common in practice.  
Table \ref{birthwtresults} shows the two most frequently appearing models of each size across solution
paths for all samples from the posterior, 
together with the relative frequency with which this model appears amongst models
with the same number of covariates.  The table only reports results for the adaptive lasso.  
The reason why we only report results for the adaptive lasso is shown in Figure \ref{hosmerloss}, 
which gives the relative loss of explanatory power as a function
of the posterior expected number of variables selected in the projection. 
In the figure, the solid line is for the lasso projection and the dashed line for
the adaptive lasso -- it can be seen that for the adaptive lasso there is a reduced
loss of explanatory power compared to the lasso for a given level of parsimony.    

\begin{figure}
\caption{\label{hosmerloss}  Plot of relative loss of explanatory power versus posterior expected 
model size for low birth weight example.  The solid line is for the lasso projection and
the dashed line is for the adaptive lasso.}
\begin{center}
\begin{tabular}{c}
\includegraphics[width=80mm,angle=90]{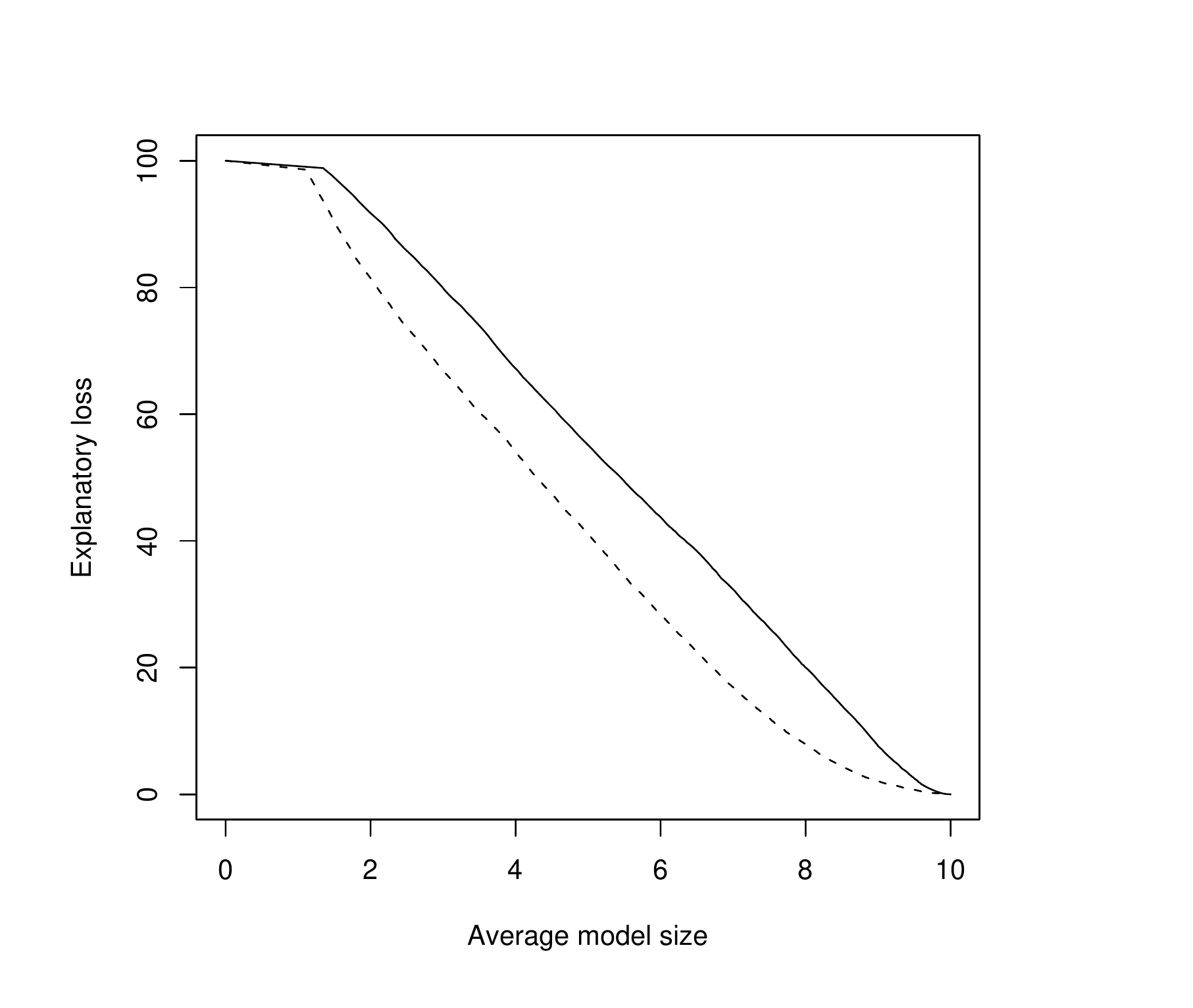} 
\end{tabular}
\end{center}
\end{figure}
Examining the models in Table \ref{birthwtresults}, the indicators for number of first trimester 
physician visits (ftv), 
one of the indicators for race and age appear to be the least important covariates.  
This is consistent with other published analyses of this data set such as in Venables and Ripley (2002).  
They consider stepwise variable selection using AIC in a main effects model including all
the covariates, which results in exclusion of the dummy variables coding for ftv and 
age.  They also consider inclusion of second order interactions and note that there is some 
evidence for an interaction between ftv and age.  Raftery and Zheng (2003) also
consider some reference Bayesian model averaging approaches to the analysis of
this dataset.  Their conclusions concerning the important variables (based on  marginal
posterior probabilities of inclusion) are similar to ours, although it should
be noted that their model is different with first trimester physician visits treated as a continuous 
covariate rather than being coded through two indicator variables as in our analysis, which follows
Venables and Ripley (2002).

\subsection{Structured variable selection example}

Our next example concerns a variable selection problem with hierarchical structure.  The data
are simulated following a similar example discussed in Yuan, Joseph and Zou (2007).  
The purpose of considering this example is to show that the method described in Section 5
which incorporates hierarchical constraints into variable selection is beneficial.  
In particular, we consider a model satisfying the strong heredity principle, and then 
show that a projection approach to variable selection which imposes strong heredity outperforms
an approach which does not impose this constraint.  By outperforms here we mean that 
for a given level of parsimony (a given value for the posterior expected number of nonzero
components of the projection) we have a greater posterior probability for the model chosen
via the projection to encompass the true model, with a relatively small loss of explanatory
power due to imposing the constraint.  

In the example of Yuan, Joseph and Zou (2007)
three predictors $X_1$, $X_2$ and $X_3$ are simulated following a multivariate normal
distribution with mean zero and $\mbox{Cov}(X_i,X_j)=\rho^{|i-j|}$ for values $\rho$ of $-0.5$, $0$ and $0.5$.  
There are $n=50$ observations simulated and $100$ different datasets are considered
for each value of $\rho$.  We consider fitting a model that includes $X_1$, $X_2$, $X_3$ 
and all second order interaction terms (nine possible terms in all - no intercept is fitted).
The true model used to generate $Y$ is
$$Y=3X_1+2X_2+1.5X_1X_2+\epsilon$$
where $\epsilon\sim N(0,9)$.  Note that this model respects the 
strong heredity principle -- for the interaction term, the corresponding main effects
are also included.  We use two variants of our non-negative garotte approach 
to fitting the data.  The first variant respects the strong heredity principle, and the
second variant does not impose any constraint.  

We considered a grid of 100 equally spaced values for $\lambda$ between $0$ and $9$ 
in the constraint $\sum_{j=1}^p \theta_j\leq \lambda$ as described in Section 5. 
Using the usual noninformative prior in the Bayesian linear model on the regression
coefficients and variance parameter of $p(\bbeta,\sigma^2)\propto \sigma^{-2}$, 
we can simulate directly from the posterior distribution without the need for
iterative methods (see, for instance, Gelman {\it et al.}, 2003).  For each simulated
dataset we generated 1000 samples from the posterior distribution. For
each value of $\lambda$ in the grid and each draw from the posterior distribution, 
we calculated projections (with and without the strong heredity constraint) 
recording the number of active variables, whether
or not the projection encompassed the true model and the Kullback-Leibler 
divergence between the full model and the projections.  

Plotting the posterior expected values of these quantities against one another for the grid of values of
$\lambda$ gives a sense of the trade off between parsimony, predictive accuracy and
identification of the important variables.  Figure \ref{strongheredity} shows 
plots of the probability of encompassing the true model and of the explanatory loss versus
posterior expected number of variables selected in the projected model for the two
projection methods (imposing strong heredity, solid line, and no constraint, broken line).  For a given level
of parsimony it can be seen that the projection method which imposes the hierarchical constraint has a higher
posterior probability of encompassing the true model so that enforcing the strong heredity principle 
when it is approporiate is helpful for obtaining more parsimonious
models and for identifying the important variables.   

\begin{figure}
\caption{\label{strongheredity}  Plots of posterior probability of encompassing the true model versus
average model size (left column) and explanatory loss versus average model size (right column). 
The parameter $\rho$ takes values of $-0.5$ (top) 
$0$ (middle) and $0.5$ (bottom). Solid line is for strong heredity and broken line no constraint.}
\begin{center}
\begin{tabular}{cc}
\includegraphics[width=55mm]{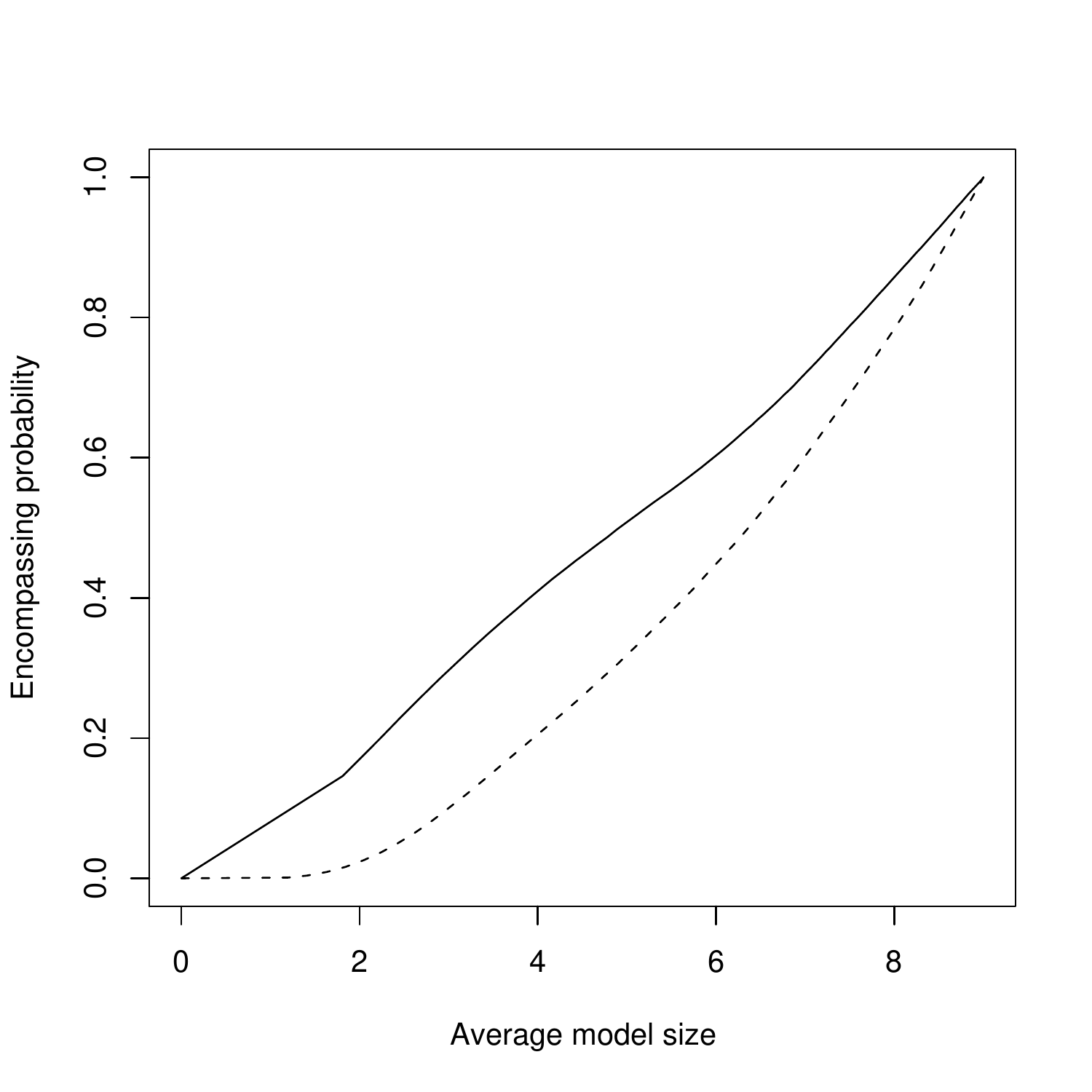} & \includegraphics[width=55mm]{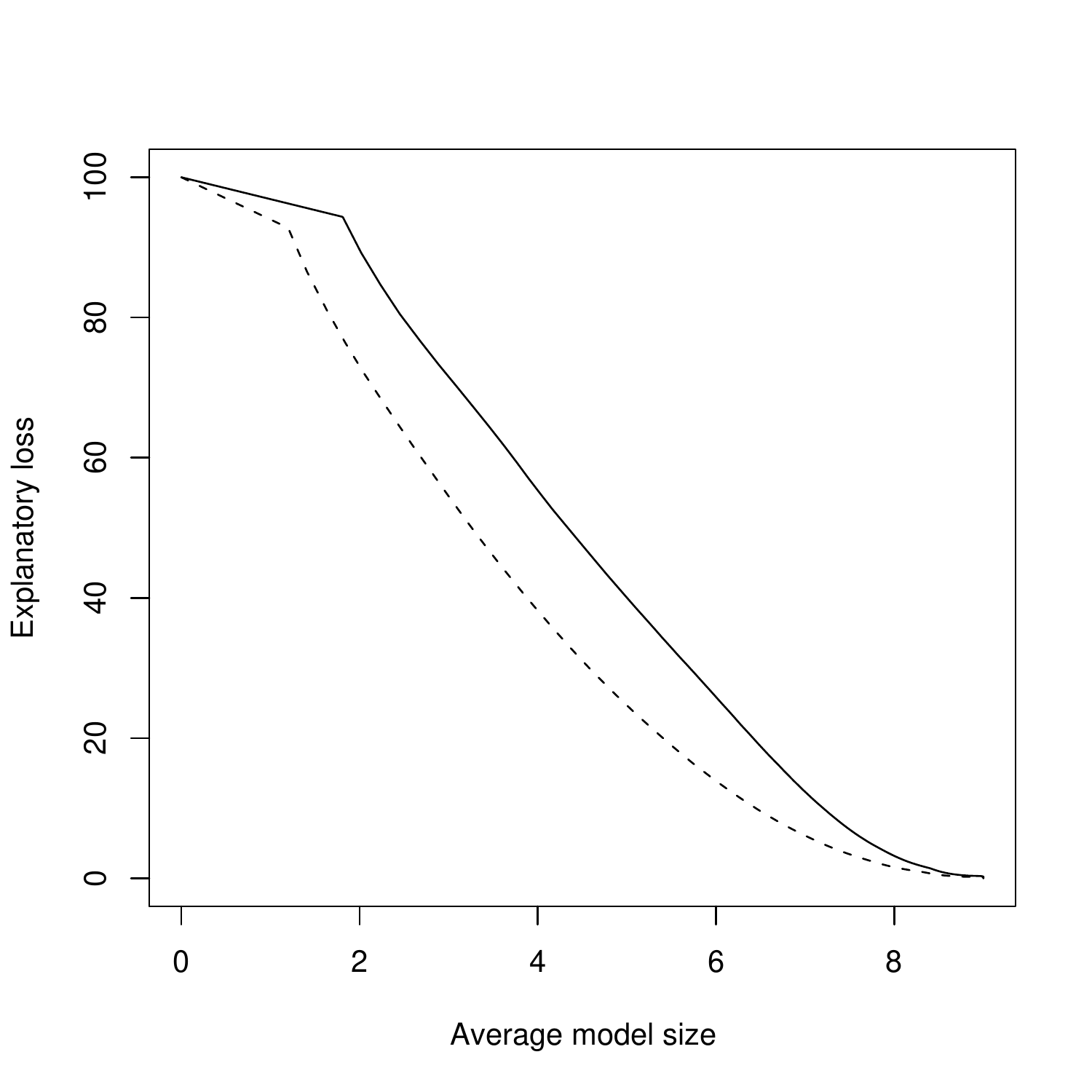} \\
\includegraphics[width=55mm]{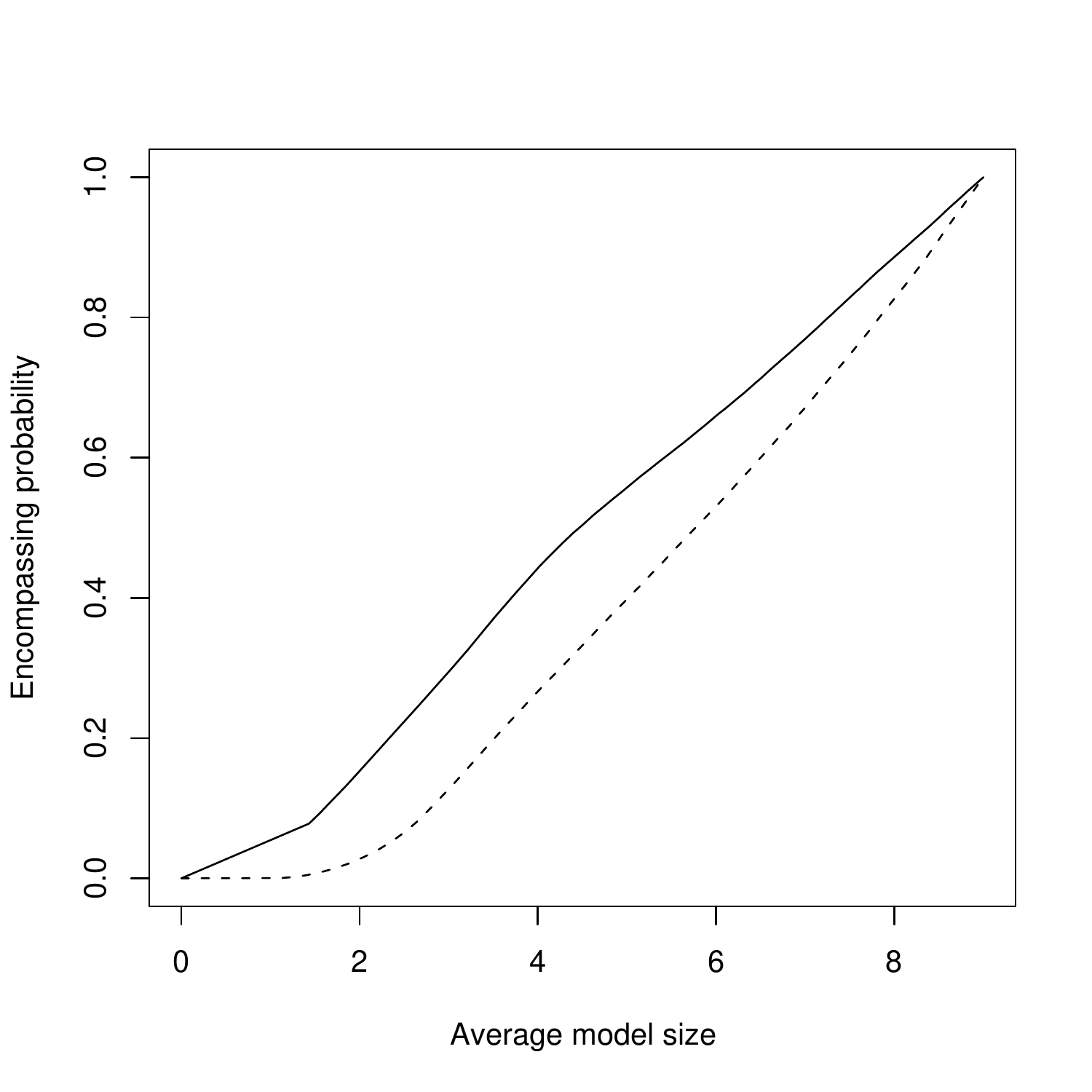}  & \includegraphics[width=55mm]{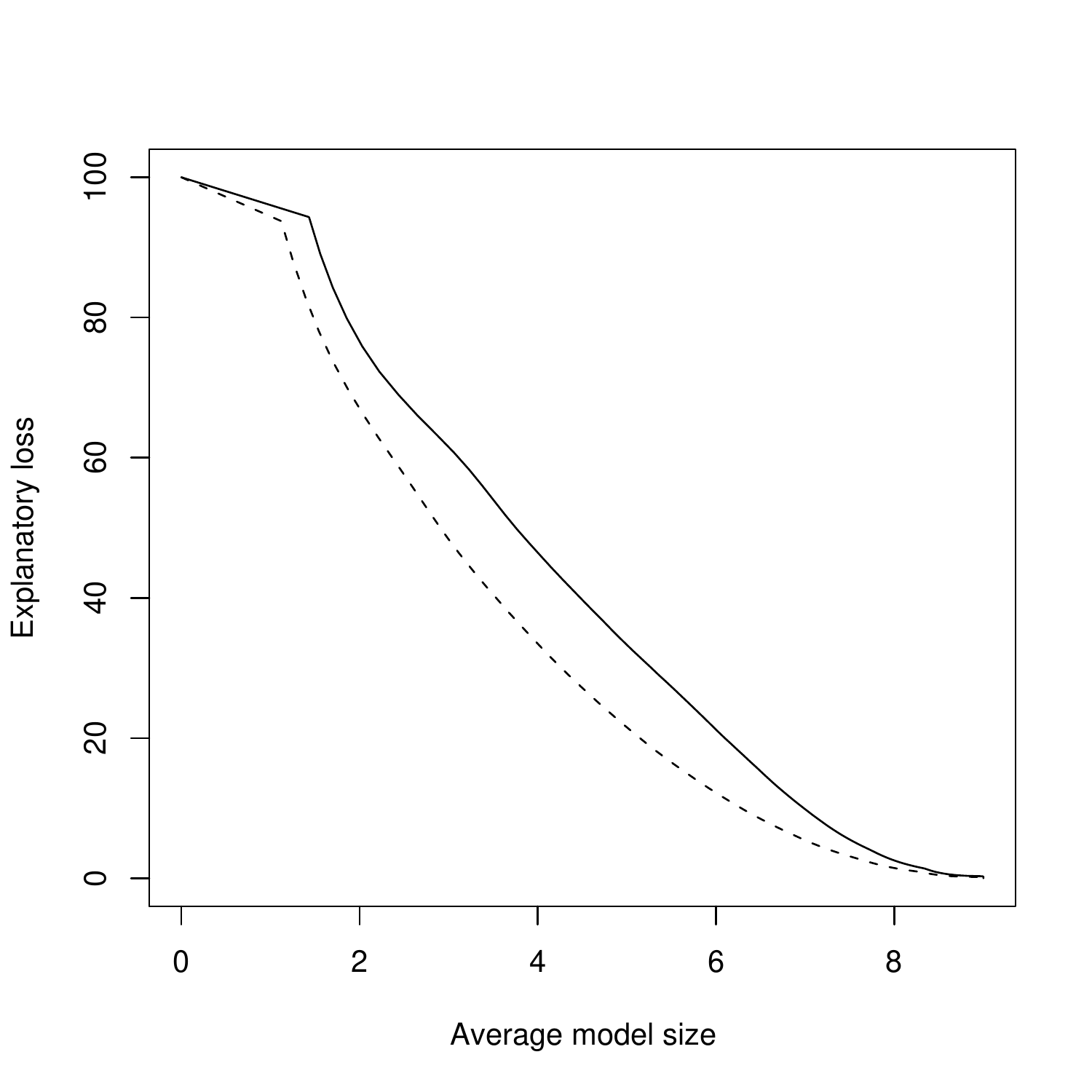} \\
\includegraphics[width=55mm]{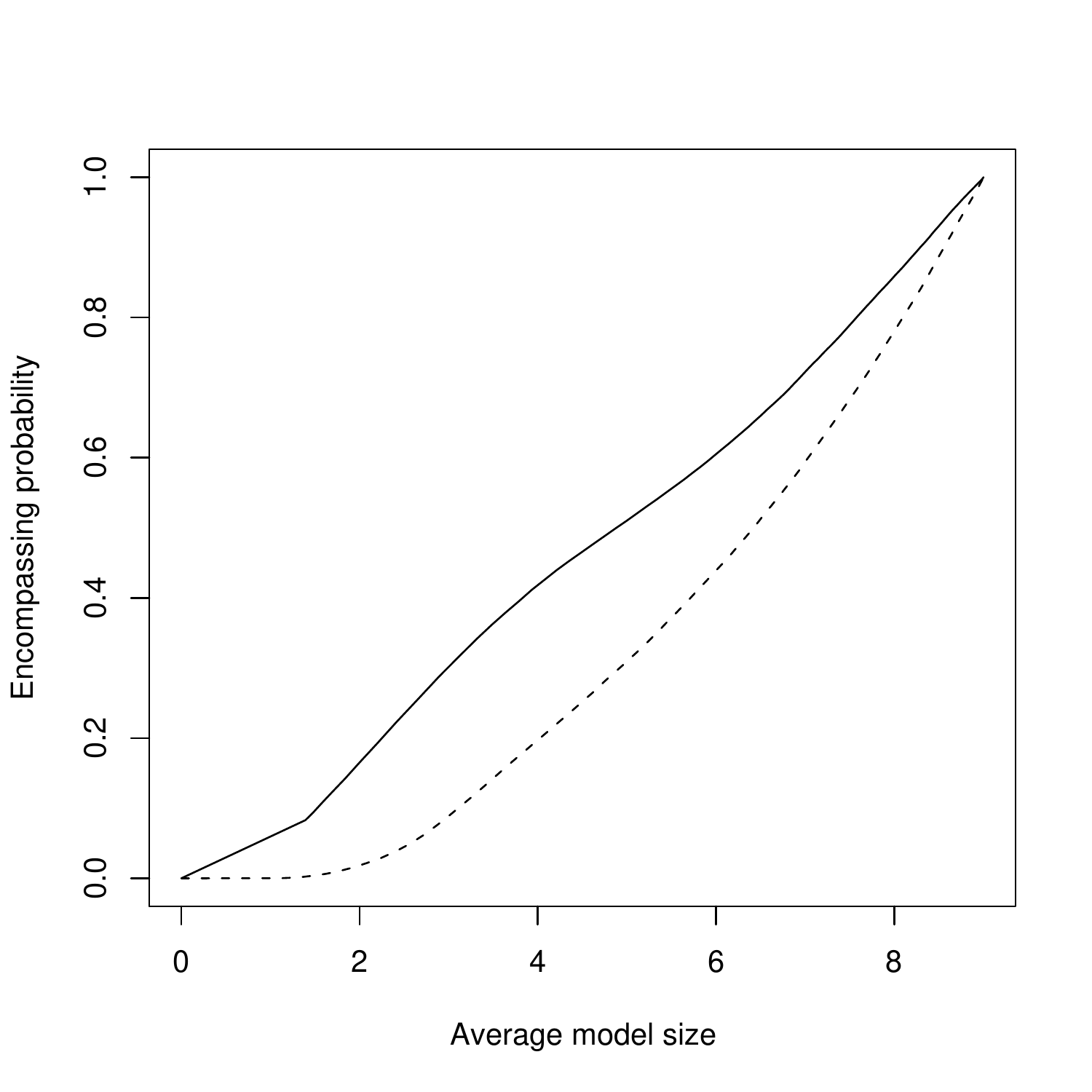} &  \includegraphics[width=55mm]{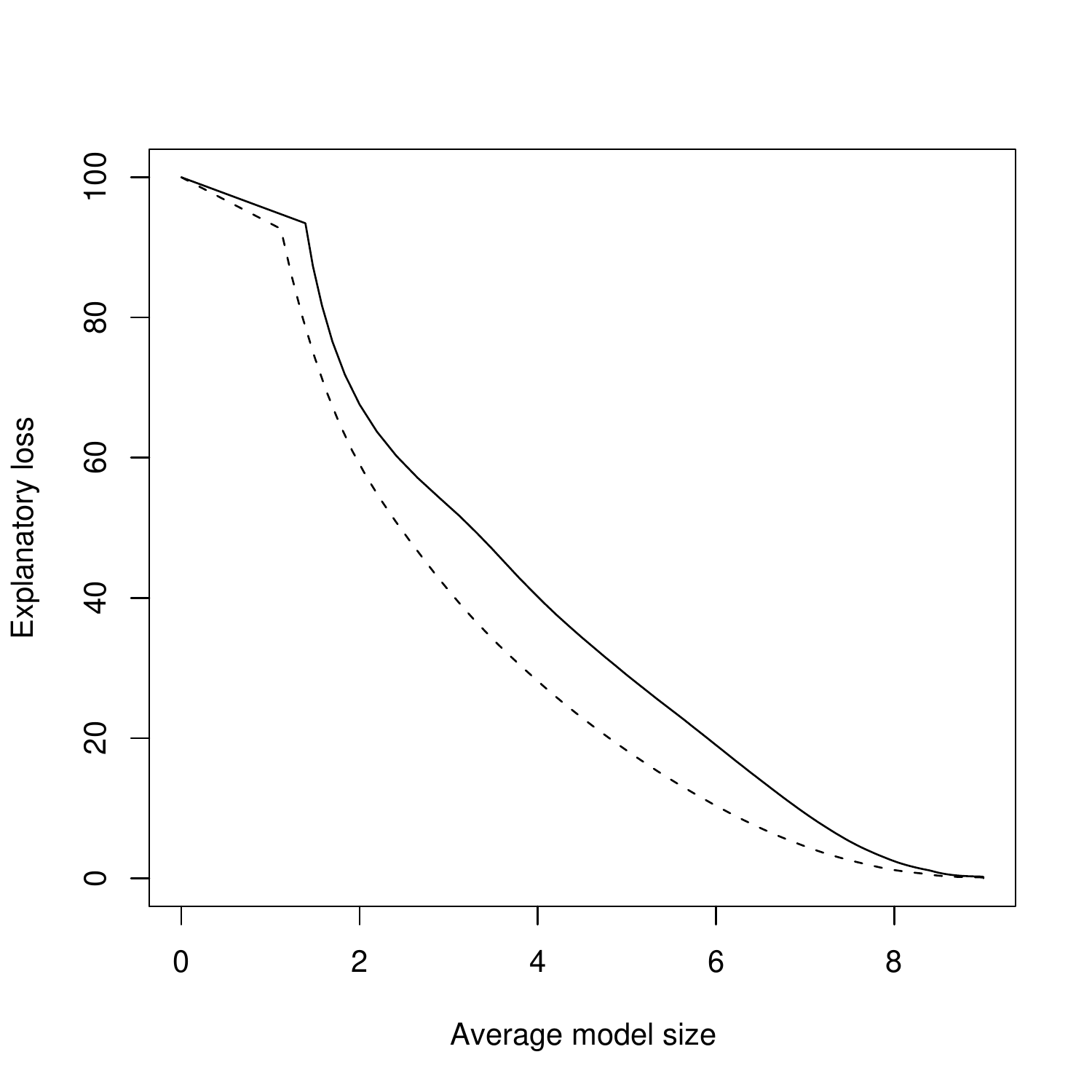}  
\end{tabular}
\end{center}
\end{figure}

\subsection{``Large $p$, small $n$" regression}

We now consider some simulations for the ``large $p$, small $n$" case where there are more
predictors than observations.  We consider generating 100 
datasets with $n=20$ and $40$ predictors.  The datasets follow a linear model
$$\by=\bX\bbeta+\bepsilon$$
where $\bepsilon\sim N(0,5^2 \bI)$.  Below we write $\bx_{i.}$ for the $i$th row of the
design matrix $\bX$.  
\begin{enumerate}  
\item Example 1: set
$\beta_j=0$, $j=1,...,10$, $j=21,...,30$, 
$\beta_j=2$, $j=11,...,20$, $j=31,...,40$.  
We have $\bx_{i.}\sim N(0,\bI)$.
\item Example 2:  set 
$\beta_j=4$, $j=1,...,5$, $\beta_j=0$, $j=6,...,40$.  
We generate $\bx_{i.}\sim N(0,\bSigma)$ with $\Sigma_{jj}=1$, $j=1,...,40$
and $\Sigma_{ij}=0.5$ $i\neq j$.  
\end{enumerate}
In the first example there is no multicollinearity, but $20$ active predictors.  
In the second example there is moderate multicollinearity but only $5$ 
active predictors.  

Since the number of predictors is double the number of observations in
both examples, here we are considering a 
``large $p$, small $n$" situation.  For a Bayesian analysis of the data with an encompassing
model we consider the Bayesian lasso of Park and Casella (2008).  
They consider the following priors on parameters.  If an intercept term $\beta_0$ is included, 
this is given a flat prior $p(\beta_0)\propto 1$ and this parameter can be integrated
out of the model analytically.  Conditional on the variance $\sigma^2$, the $\beta_j$ are conditionally
independent in their prior with 
$$p(\beta_j|\sigma^2)=\frac{\lambda}{2\sigma^2}\exp\left(-\frac{\lambda |\beta_j|}{\sqrt{\sigma^2}}\right)$$
where $\lambda>0$ is a shrinkage parameter.  Finally an inverse gamma prior can be used
for $\sigma^2$, where we use $IG(0.01,0.01)$.  
A hyperprior can be placed on $\lambda$, or it can be estimated by marginal maximum likelihood as outlined
in Park and Casella (2008) or by cross-validation.  For illustrative purposes here we will
fix $\lambda=10$ in the computations below.  Park and Casella (2008) outline an efficient MCMC
scheme for computations.  

We consider projections based
on the adaptive lasso, and Figure \ref{lpsn} shows a plot of the false discovery rate 
(average number of variables incorrectly selected divided by average number of variables
selected) versus average model size.  The averages are over $100$ simulation replicates.  
Quite a large model would need to be chosen to encompass all the active
predictors.  Note that if we use the classical lasso to do selection then the number of predictors
chosen by the projection cannot be more than the number of observations.  Figure \ref{lpsn} also
shows the explanatory loss as a function of the average model size.  Model uncertainty
is considerable here, and we believe that the distribution on the model space defined by the 
projection is extremely valuable for exploring model uncertainty.  Figure \ref{lpsn2} shows for the
first simulation replicate in each example the marginal posterior probabilities of the variables being 
nonzero in the projection.  The projections in the figure correspond to average model size
of $13$ (example 1) and $10$ (example 2) corresponding to approximately $20$\% 
explanatory loss in both cases.  The lines show the mean values for these probabilities within the
active and inactive groups.  It is clear that there is some useful information in the posterior
distribution of the projection for distinguishing active from inactive variables.  
It is important to realize that Figure \ref{lpsn2} is examining posterior probabilities of selection
in the projection for a single replicate, not the frequentist behaviour of selection across replicates - such
frequentist behaviour is summarized by the false discovery rates of Figure \ref{lpsn}.  We also
stress that posterior probabilities of selection in the projection depend on the prior in the
encompassing model and the tolerable explanatory loss.  
\begin{figure}
\caption{\label{lpsn}  Plots of false discovery rate versus
average model size (top row) and explanatory loss versus average model size (bottom row) for
examples 1 (left) and 2 (right). Plotted points correspond
to a grid of values for the constraint $\lambda$.}
\begin{center}
\begin{tabular}{ccc}
\includegraphics[width=65mm]{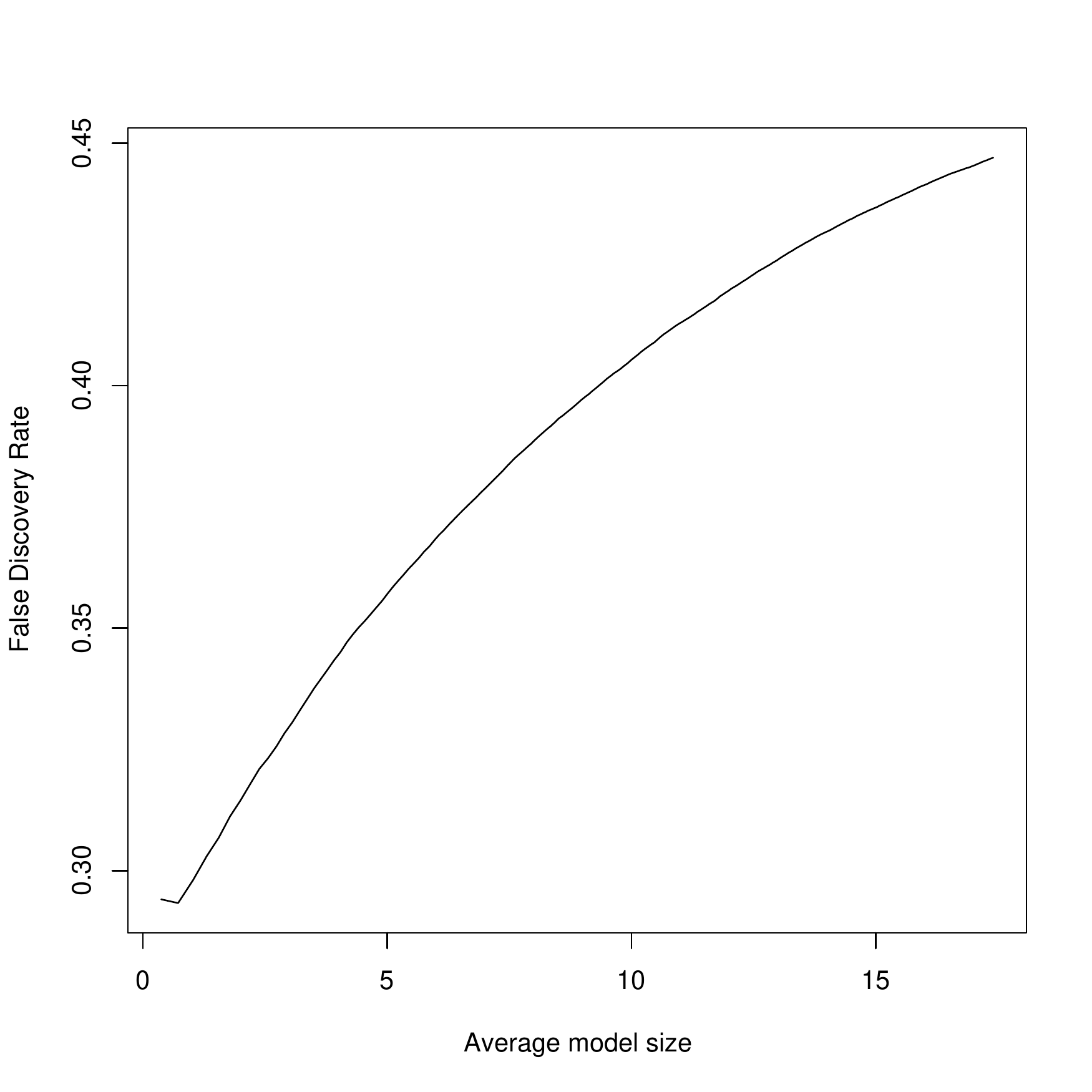} & \includegraphics[width=65mm]{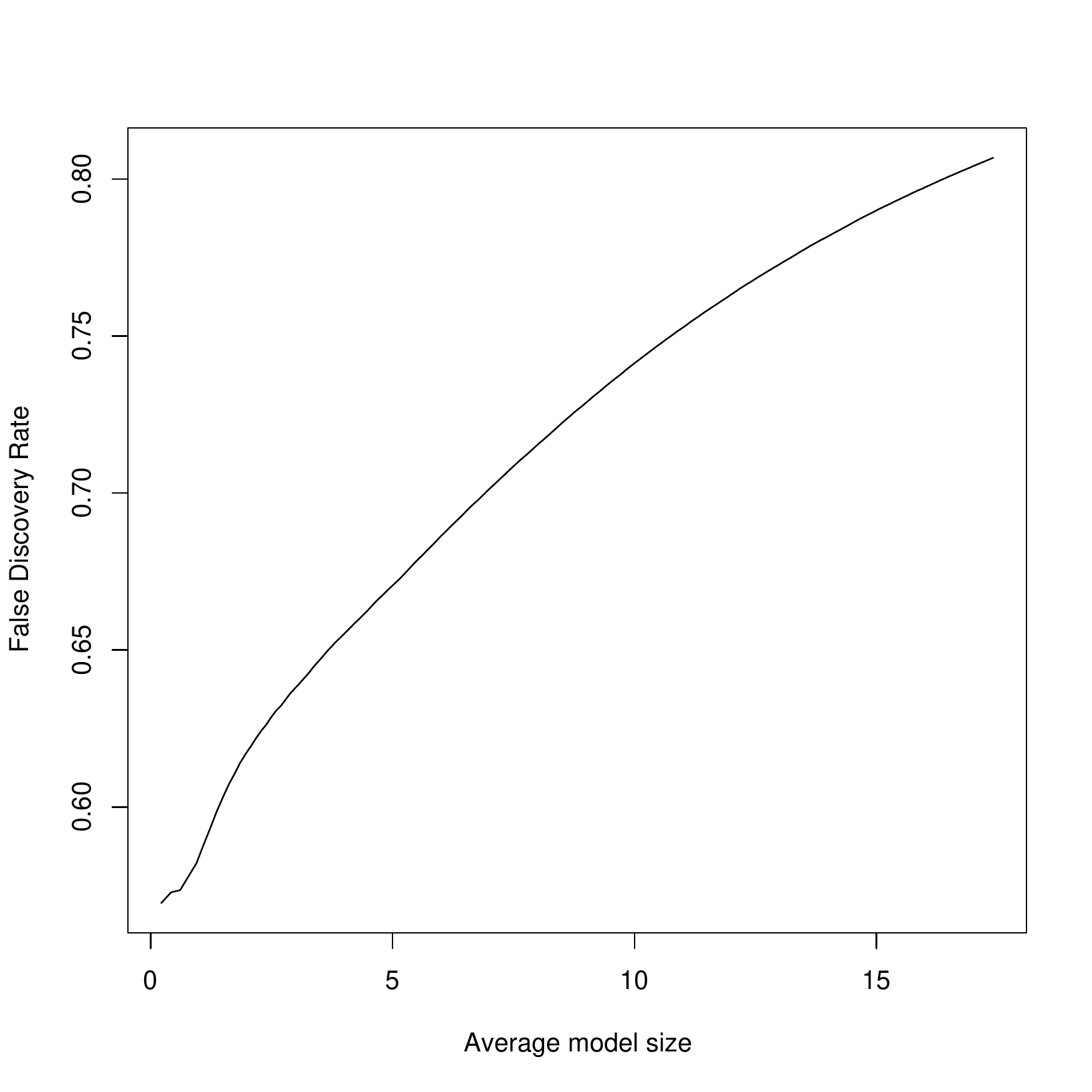}   \\
\includegraphics[width=65mm]{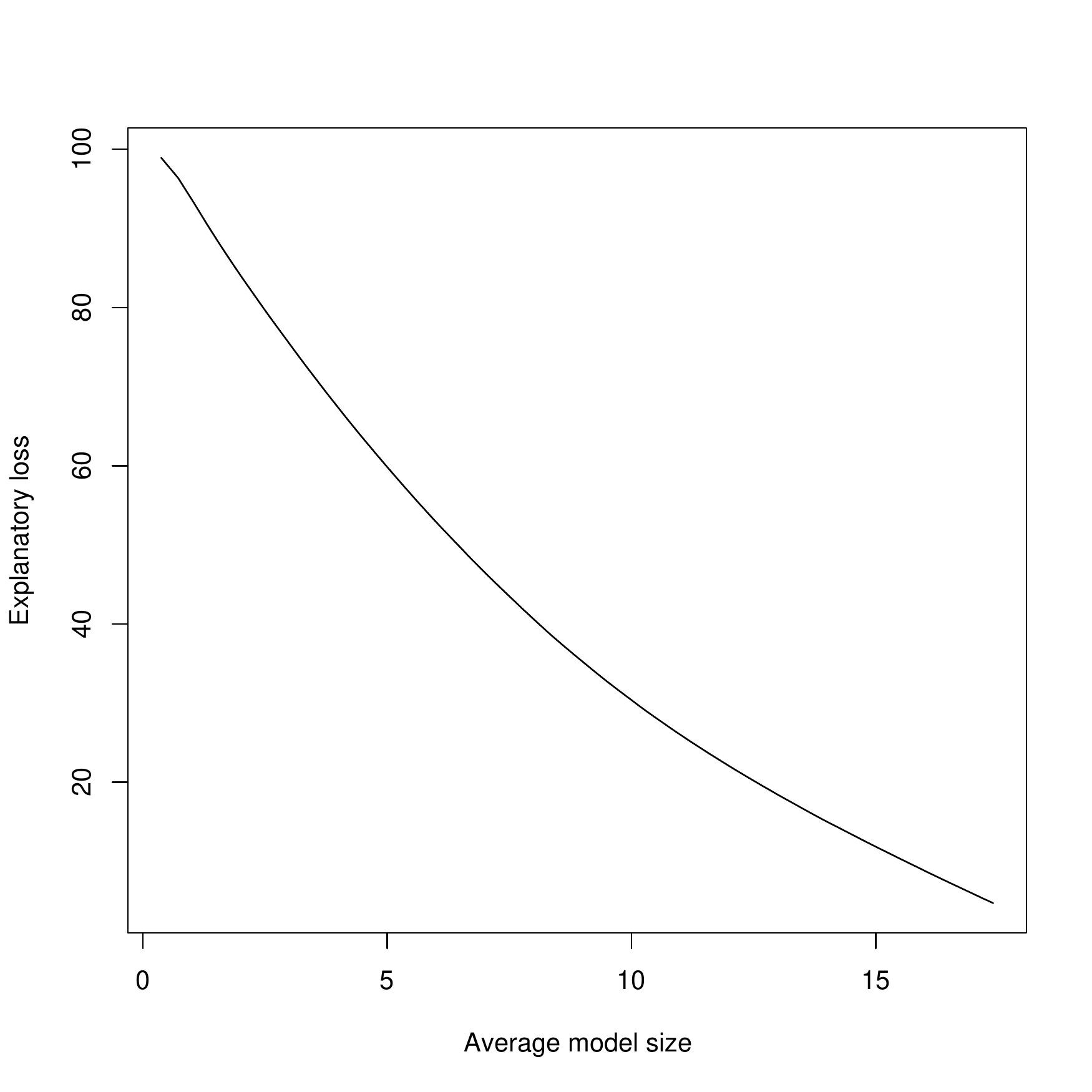}  & \includegraphics[width=65mm]{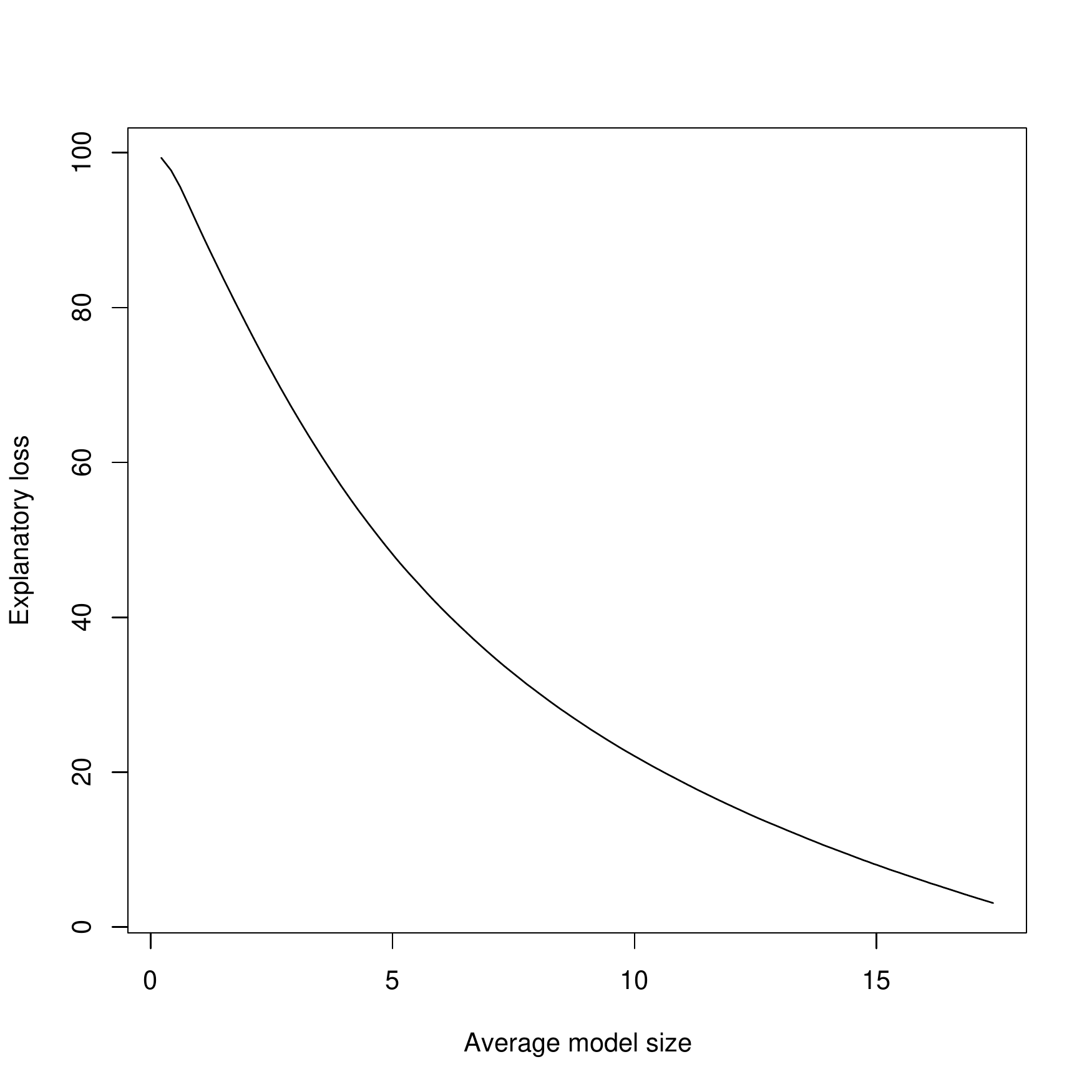}  
\end{tabular}
\end{center}
\end{figure}

\begin{figure}
\caption{\label{lpsn2}  Plots of marginal posterior probability of inclusion versus variable for
first simulation replicate for example 1 (left) and 2 (right). Probabilities are for projections with
average model size 13 (left) and 10 (right) corresponding in both cases to approximately 
$20$\% explanatory loss.  The lines
show the mean posterior probabilities of inclusion among the active and inactive groups.}
\begin{center}
\begin{tabular}{ccc}
\includegraphics[width=65mm]{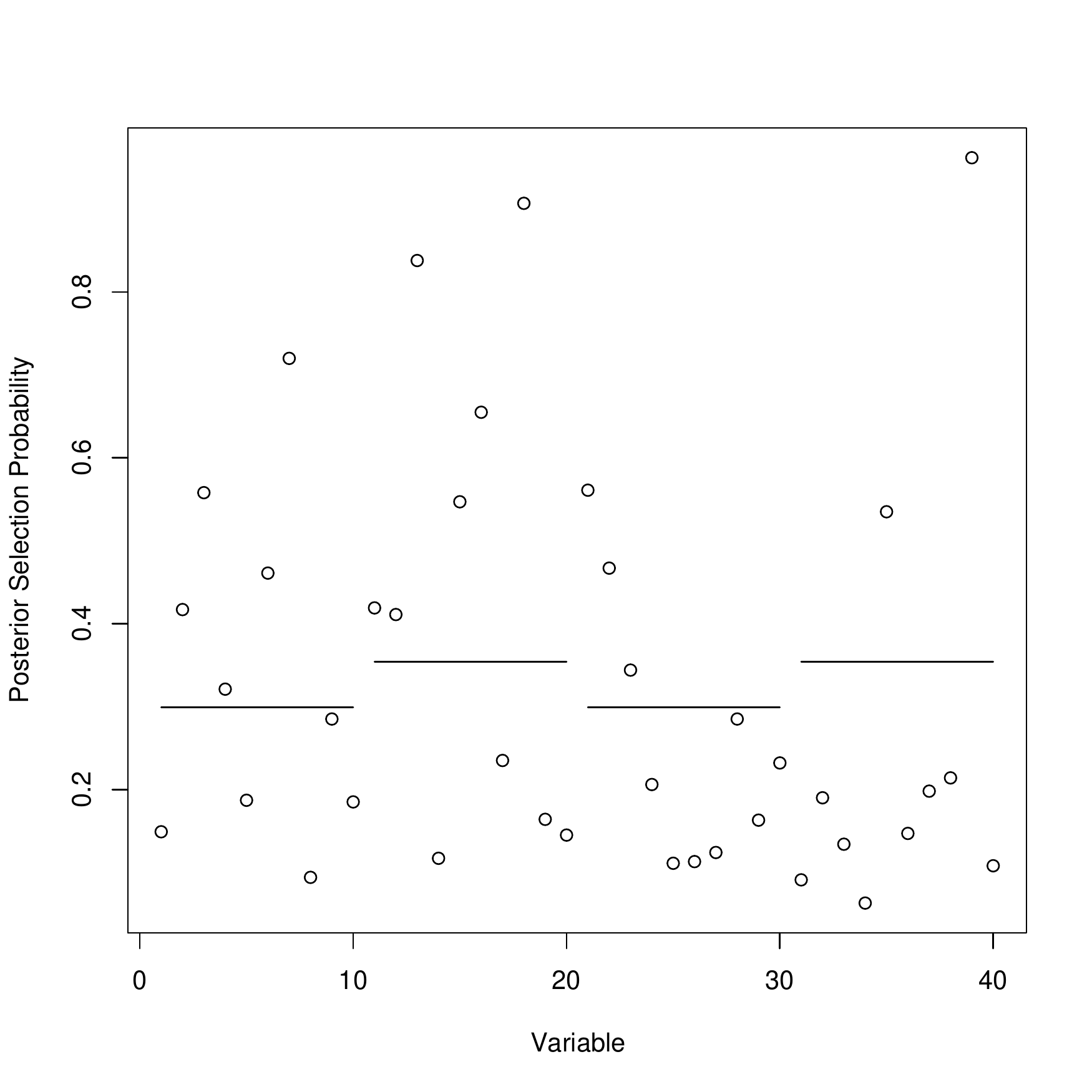} & \includegraphics[width=65mm]{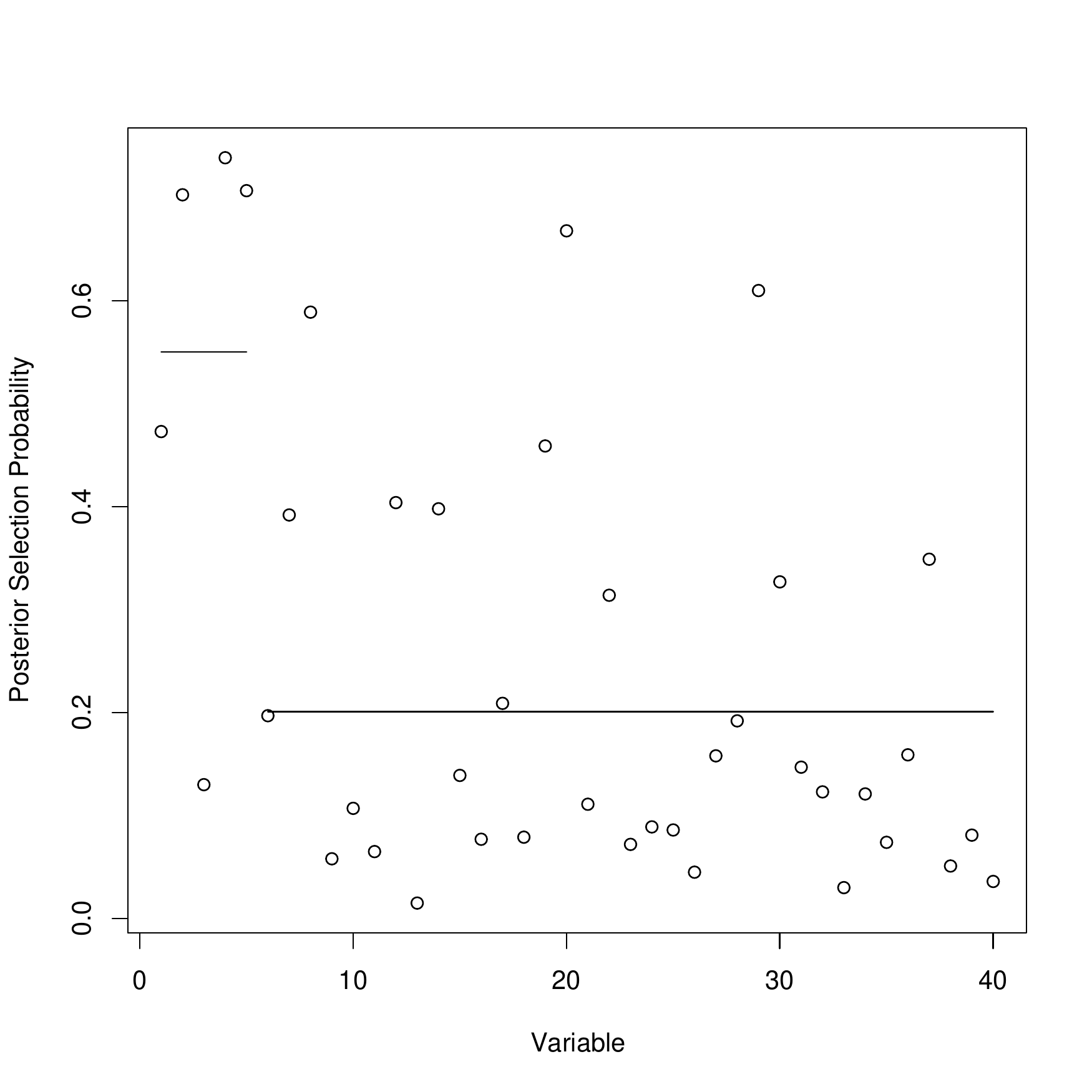}   
\end{tabular}
\end{center}
\end{figure}

\section{Conclusion}

We have discussed the use of Kullback-Leibler projections related to the lasso as a tool for the
exploration of model uncertainty.  There are many possible extensions to our suggested
framework.  One interesting possibility which we are currently pursuing is the use of projections
related to versions of the lasso for selection on batches of parameters and random effects.  

\section*{Appendix}

\noindent
{\em Proof of Theorem 1}:
We write $\bbeta=\bbeta^0+\bv/\sn$, where $\|\bv\|\le C$.
Denoting $\bu=\sqrt{n}(\bbeta_S-\bbeta^0)$, we define
\[ L(\bu)= \sum_{i=1}^n \{-\mu_i(\bbeta) \bx_i^T(\bbeta^0+\frac{\bu}{\sn})+
b(\bx_i^T(\bbeta^0+\frac{\bu}{\sn}))\}
+\gamma \sum_{j=1}^p |\beta_{j}^0+\frac{u_j}{\sn}|/|\beta_j|\]
and $Z(\bu)=L(\bu)-L({\bf 0}).$  The minimizer $\bu'$ of $L(\bu)$ gives the minimizer
$\bbeta_S'$ of (\ref{obj2}) by $\bbeta_S'=\bbeta^0+\bu'/\sqrt{n}$,  and hence to
study $\bbeta_S'$ it suffices to consider $\bu'$.  Following Zou (2006), we decompose $Z(\bu)$ as
\[ Z(\bu)= Z_1(\bu)+Z_2(\bu)+Z_3(\bu)+Z_4(\bu)\]
where
\begin{align*}
  Z_1(\bu)&= -\sum_{i=1}^n [\mu_i(\bbeta)- b'(\bx_i^T\bbeta^0)]
  \frac{\bx_i^T \bu}{\sn}\\
  Z_2(\bu)& = \sum_{i=1}^n \frac{1}{2} b''(\bx_i^T\bbeta^0) \bu^T \frac{\bx_i \bx_i^T}
  {n} \bu\\
  Z_3(\bu) &= \gamma \sum_{j=1}^p \frac{|\beta_{j}^0+\frac{u_j}{\sn}|-|\beta_j^0|}
  {|\beta_j|}\\
  Z_4(\bu) &= n^{-3/2} \sum_{i=1}^n \frac{1}{6} b'''(\bx_i^T \bbeta^*)(\bx_i^T\bu)^3
\end{align*}
where $\bbeta^*$ lies between $\bbeta^0$ and $\bbeta^0+\bu/\sn$.
Since $\mu_i(\bbeta^0)=b'(\bbeta^0)$, we can write the first term as
\[ Z_1(\bu)=\sum_{i=1}^n (\mu_i(\bbeta)-\mu_i(\bbeta^0))\frac{\bx_i^T \bu}{\sn}=
\sum_{i=1}^n(\frac{\bv^T}{\sn}+o_p(1/\sn)) \mu_i'(\bbeta^0) \frac{\bx_i^T \bu}{\sn}
=\ba_n^T \bu+o_p(1),
\]
where $\| \ba_n \| =O_p(1).$
For the second term $Z_2(\bu)$, we have
\[ \sum_{i=1}^n b''(\bx_i^T\bbeta^0) \frac{\bx_i \bx_i^T}{n} \rightarrow I(\bbeta^0). \]
Thus $Z_2(\bu) \rightarrow 1/2 ~\bu^T I(\bbeta^0) \bu$. For the third term, 
following the arguments in Zou (2006), we have
\begin{equation*}
  \gamma_n  \frac{|\beta_{j}^0+\frac{u_j}{\sn}|-|\beta_j^0|}
  {|\beta_j|}
  \rightarrow_p 
  \begin{cases}
    0 & \beta^0_j \ne 0\\
    0 & \beta^0_j=0 ~\text{and}~ u_j=0\\
    \infty & \beta^0_j = 0 ~\text{and}~ u_j \ne 0
  \end{cases}
\end{equation*}
since $\gamma_n$ satisfies $\gamma_n/\sn \rightarrow 0$ and $\gamma_n \rightarrow 
\infty$.
The fourth term is of the
order $O_p(1/\sn)$ as
\[ 6 \sn Z_4(\bu) \le \sum_{i=1}^n \frac{1}{n}M(x_i)|\bx_i^T\bu|^3 \rightarrow_p 
E[M(\bx)|\bx^T\bu|^3] < \infty. \]
From the above arguments, we must have
\[ \bu_\mA =  O_p(1)~\text{and}~\bu_{\mA^C} \rightarrow_d 0, \]
where $\bu_\mA$ is the subvector of $\bu$ corresponding to the coefficients in 
$\mA(\bbeta^0)$. Thus $\bbeta_S'$ is $\sn$-consistent, and  
$\forall j \in \mA(\bbeta^0)$, with probability tending to one, $\bbeta^0_j$ is
estimated by a nonzero coefficient. 

It suffices then to show that
$\forall k \notin \mA(\bbeta^0)$, with probability tending to one, $\beta^0_{k}$ will
be estimated by zero. Otherwise, by the Karush-Kuhn-Tucker optimality conditions, we must
have
\begin{equation}
 \frac{1}{\sn}\sum_{i=1}^n \bx_{ik}(\mu_i(\bbeta)-b'(\bx_i^T \bbeta_S))
=\frac{\gamma_n}{\sn |\beta_{k}|} \text{sgn}(\beta_{S,k}). \label{eq1} 
\end{equation}
It is easy to see that the left hand side is equivalent to 
\begin{align*}
&\frac{1}{\sn}\sum_{i=1}^n x_{i k}[(\mu_i(\bbeta)-\mu_i(\bbeta^0))
-(b'(\bx_i^T \bbeta_S)-b'(\bx_i^T\bbeta^0))]\\
=&\frac{1}{\sn}\sum_{i=1}^n x_{ik}
[(\bbeta-\bbeta^0)^T\mu_i'(\bbeta^0)
-b''(\bx_i^T \bbeta^0) \bx_i^T (\bbeta_S-\bbeta^0)+o_p(\| \bbeta-\bbeta^0\|
+\|\bbeta_S-\bbeta^0 \|)]\\
=&O_p(1).
\end{align*}
However, the right hand side satisfies
\[ \frac{\gamma_n}{\sn |\beta_{k}|} \rightarrow \infty, \]
since $\beta_k = \beta_k^0+O_p(1/\sn)=O_p(1/\sn)$ for $\beta_k \in \mN$
and $\beta_k \notin \mA$.
This contradicts (\ref{eq1}) and the proof is completed.

\section*{References}

\bib
Berger, J. and Pericchi, L. (1996) 
The intrinsic Bayes factor for model selection and prediction.
{\it J. Amer. Statist. Assoc.}, 91, 109-122. 

\bib
Bernardo, J. M. and Rueda, R. (2002). Bayesian hypothesis testing: A reference approach. 
{\it Int. Statist. Rev.}, 70, 351-372. 

\bib
Breiman, L. (1995) Better subset regression using the non-negative garotte. 
{\it Technometrics}, 3, 373-384.

\bib
Brown, P.J., Fearn, T. and Vannucci, M. (1999)
The choice of variables in multivariate regression:  A
non-conjugate Bayesian decision theory approach.
{\it Biometrika}, 86, 635--648.

\bib
Brown, P.J., Vannucci, M. and Fearn, T. (2002)
Bayes Model averaging with selection of regressors.  
{\it J. Roy. Statist. Soc. B}, 64, 519--536.

\bib
Chipman, H. (1996).  
Bayesian variable selection with related predictors.  
{\it Canadian J. Statist.}, 24, 17--36.

\bib
Draper, D. and Fouskakis, D. (2000)
A case study of stochastic optimization in health policy:  problem
formulation and preliminary results.  {\it J. Global Optimizn.}, 
18, 399--416.  

\bib
Dupuis, J.A. and Robert, C.P. (2003)
Variable selection in qualitative models via an entropic 
explanatory power.  {\it J. Statist. Plan. Inf.}, 
111, 77--94.  

\bib
Fern\'{a}ndez, C., Ley, E., and Steel, M.F.J. (2001) 
Benchmark priors for Bayesian model averaging.
{\it J. Economet.}, 100, 381--427.  

\bib
Gelfand, A.E. and Ghosh, S.K. (1998)  
Model choice:  a minimum posterior predictive loss
approach.  {\it Biometrika}, 85, 1--11.  

\bib
Gelman, A., Carlin, J. B., Stern, H. S., and Rubin, D. B. (2003). 
Bayesian Data. Analysis (2nd edition). London: CRC Press.

\bib
Gelman, A., Meng, X.-L. and Stern, H. (1996).  
Posterior predictive assessment of model fitness via realized
discrepancies.  {\it Statistica Sinica}, 6, 733--807. 

\bib
George, E.I. and Foster, D.P. (2000)  
Calibration and empirical
Bayes variable selection. {\it Biometrika}, 87, 731--747.

\bib
Griffin, J.E. and Brown, P.J. (2007).  
Bayesian adaptive lassos with non-convex penalization.  
Technical report available at \\
\verb+http://www.kent.ac.uk/ims/personal/jeg28/BALasso.pdf+

\bib
Goutis, C. and Robert, C.P. (1998).  
Model choice in generalised linear models: A Bayesian approach via Kullback-Leibler projections. 
{\it Biometrika}, 85, 29-37

\bib
Hoeting, J.A., Madigan, D., Raftery, A.E. and Volinsky, C.T. (1999). 
Bayesian model averaging: A tutorial (with Discussion). 
{\it Statistical Science}, 14, 382--401. 
Correction: vol. 15, pp. 193-195. Corrected version available at \\ \verb+http://www.stat.washington.edu/www/research/online/hoeting1999.pdf+

\bib
Kohn, R., Smith, M. and Chan, D. (2001).  
Nonparametric regression using linear combinations of basis functions.  
{\it Statistics and Computing}, 11, 313--322. 

\bib
Lindley, D.V. (1968)
The choice of variables in multiple regression (with discussion).
{\it J. Roy. Statist. Soc. B}, 30, 31--66.

\bib
Martin, A. and Quinn, M. (2007).  
The MCMCpack package (version 0.9-1).  R package manual available at \\
\verb+http://cran.r-project.org/doc/packages/MCMCpack.pdf+

\bib
Mengersen, K. and Robert, C. (1996). Testing for mixtures: aBayesian entropy approach. 
In: Bayesian Statistics 5,  Eds. J.O. Berger, J.M. Bernardo, A.P. Dawid, D.V. Lindley and 
A.F.M. Smith. pp. 255--276, Oxford University Press.

\bib
O'Hagan, A. (1995).  
Fractional Bayes factors for model comparison (with discussion).  
{\it J. Roy. Statist. Soc. B}, 56, 99--138.  

\bib
Osborne, M.R., Presnell, B. and Turlach, B.A. (2000). 
A new approach to variable selection in least squares problems. 
{\it IMA Journal of Numerical Analysis}, 20, 389-403. 

\bib
Park, T. and Casella, G. (2008).  The Bayesian lasso.  
{\it J. Amer. Statist. Assoc.}, 103, 681--686.

\bib
Park, M.-Y. and Hastie, T. (2007). 
An L1 regularization-path algorithm for generalized linear models. 
{\it J. Roy. Statist. Soc. B}, 69, 659–677. 

\bib
Paul, D., Bair, E., Hastie, T. and Tibshirani, R. (2007) 
Pre-conditioning for feature selection and regression in high-dimensional problems. 
{\it Annals of Statistics}, to appear.  

\bib
Raftery, A.E. (1996).  
Approximate Bayes factors and accounting for model uncertainty in generalized
linear models.  {\it Biometrika}, 83, 251-266.  

\bib
Raftery, A.E. and Zheng, Y. (2003).  
Discussion:  Performance of Bayesian Model Averaging.   
{\it J. Amer. Statist. Assoc.}, 98, 931--938.  

\bib
Spiegelhalter, D.J., Best, N.G., Carlin, B.P. and van der Linde, A. 
(2002)  Bayesian measures of model complexity and fit (with discussion).  
{\it J. Roy. Statist. Soc. B}, 64, 583--639.  

\bib
Tibshirani, R. (1996) 
Regression shrinkage and selection via the lasso. 
{\it J. Roy. Statist. Soc. B}, 58, 267--88. 

\bib
Vehtari, A. and Lampinen, J. (2004). 
Model Selection via Predictive Explanatory Power. 
Report B38, Laboratory of Computational Engineering, Helsinki University of Technology. 

\bib
Yuan, M., Joseph, V.R. and Zou, H. (2007).  
Structured variable selection and estimation.  
Technical report.  Available at \\
\verb+http://www2.isye.gatech.edu/~myuan/YuanPub.html+

\bib
Yuan, M. and Lin, Y. (2005).  
Efficient empirical Bayes variable selection and estimation.  
{\it J. Amer. Statist. Assoc.}, 100, 1215--1225.  

\bib
Zou, H. and Hastie, T. (2005)  
Regularization and variable selection via the elastic net.  
{\it J. Roy. Statist. Soc. B}, 67, 301--320.

\bib
Zou, H. (2006).  
The adaptive lasso and its oracle properties.  
{\it J. Amer. Statist. Assoc.}, 101, 1418--1429.  

\newpage

\begin{table}[c]
\caption{\label{birthwtpred}Predictors for low birth weights data set}
\begin{center}
\begin{tabular}{rl}
\hline \hline 
Predictor & \multicolumn{1}{c}{Description} \\ \hline
age & age of mother in years \\
lwt & weight of mother (lbs) at least menstrual period \\
raceblack & indicator for race=black (0/1) \\
raceother & indicator for race other than white or black (0/1) \\
smoke & smoking status during pregnancy (0/1) \\
ptd & previous premature labors (0/1) \\
ht & history of hypertension (0/1) \\
ui & has uterine irritability (0/1) \\
ftv1 & indicator for one physician visit in first trimester (0/1) \\
ftv2+ & indicator for two or more physician visits in first trimester (0/1) \\ \hline
\end{tabular}
\end{center}
\end{table}

\begin{table}[c]
\caption{\label{birthwt} Posterior means and standard deviations of
coefficients for full model fitted to low
birthweight data.}
\begin{center}
\begin{tabular}{cll}
\hline \hline
Predictor & Posterior & Posterior \\
              & Mean      & Standard \\
             &                 & Deviation \\ \hline
age        & -0.21  &  0.21  \\
lwt         & -0.48  &  0.22  \\ 
raceblack &   1.06 &  0.52 \\ 
raceother  &  0.65  &  0.43 \\  
smoke     &   0.68  &  0.41 \\  
ptd          &   1.31  &  0.47 \\  
ht            &   1.69  &  0.67 \\  
ui           &  0.64  & 0.46 \\  
ftv1         & -0.49  & 0.46 \\  
ftv2         & 0.11   & 0.44  \\ \hline
\end{tabular}
\end{center}
\end{table}

\begin{table}[c]
\caption{\label{birthwtresults} Two most frequently appearing models of each size in 
solution path for the projection together with relative frequency of each model within
all appearances of model of the same size (Prob/Size).  
 Zeros and ones in the columns labelled by the predictors show inclusion and exclusion for 
different models (rows). }
\begin{center}
\begin{tabular}{crrrrrrrrrrr}
\hline \hline
Model & \multicolumn{10}{c}{Predictor} & Prob/ \\ 
Size   &       &      &         &         &             &       &    &    &      &       & Size \\ \hline
         & age & lwt & black & other & smoke & ptd & ht & ui & ftv1& ftv2  &  \\ 
 1  &  0 &   0 &   0 &   0 &   0 &   1 &   0 &   0 &   0 &    0 &  0.48 \\
 1  &  0 &   1 &   0 &   0 &   0 &   0 &   0 &   0 &   0 &    0 &  0.17 \\
 2  &  0 &   1 &   0 &   0 &   0 &   1 &   0 &   0 &   0 &    0 &  0.24 \\
 2  &  0 &   0 &   0 &   0 &   0 &   1 &   1 &   0 &   0 &    0 &  0.10 \\
 3  &  0 &   1 &   0 &   0 &   0 &   1 &   1 &   0 &   0 &    0 &  0.13 \\
 3  &  0 &   1 &   1 &   0 &   0 &   1 &   0 &   0 &   0 &    0 &  0.06 \\
 4  &  0 &   1 &   0 &   0 &   1 &   1 &   1 &   0 &   0 &    0 &  0.07 \\
 4  &  0 &   1 &   0 &   0 &   0 &   1 &   1 &   0 &   1 &    0 &  0.06 \\
 5  &  0 &   1 &   1 &   0 &   1 &   1 &   1 &   0 &   0 &    0 &  0.05 \\
 5  &  0 &   1 &   1 &   0 &   0 &   1 &   1 &   1 &   0 &    0 &  0.05 \\
 6  &  0 &   1 &   1 &   1 &   1 &   1 &   1 &   0 &   0 &    0 &  0.06 \\
 6  &  1 &   1 &   1 &   0 &   1 &   1 &   1 &   0 &   0 &    0 &  0.05 \\
 7  &  0 &   1 &   1 &   1 &   1 &   1 &   1 &   1 &   0 &    0 &  0.10 \\
 7  &  1 &   1 &   1 &   1 &   1 &   1 &   1 &   0 &   0 &    0 &  0.09 \\
 8  &  1 &   1 &   1 &   1 &   1 &   1 &   1 &   1 &   0 &    0 &  0.13 \\
 8  &  0 &   1 &   1 &   1 &   1 &   1 &   1 &   1 &   1 &    0 &  0.13 \\
 9  &  1 &   1 &   1 &   1 &   1 &   1 &   1 &   1 &   1 &    0 &  0.29 \\
 9  &  1 &   1 &   1 &   1 &   1 &   1 &   1 &   1 &   0 &    1 &  0.19 \\ \hline
\end{tabular}
\end{center}
\end{table}

\end{document}